\documentclass[journal=jcctee,manuscript=article]{achemso}
\usepackage{psfrag} \usepackage{stdclsdv}
\usepackage{rotating}
\usepackage{xcolor}
\usepackage{mathtools}
\usepackage{bm}
\setkeys{acs}{maxauthors=50}
\usepackage{amsmath}
\setkeys{acs}{usetitle = true}
\author{Debayan Chakraborty}
\affiliation{Department of Chemistry, The University of Texas at Austin, 
Austin TX 78712, USA}
\author{Naoto Hori}
\affiliation{Department of Chemistry, The University of Texas at Austin,
Austin TX 78712, USA}
\author{D. Thirumalai}
\email{dave.thirumalai@gmail.com}
\affiliation{Department of Chemistry, The University of Texas at Austin,
Austin TX 78712, USA}
\title
{Sequence-dependent Three Interaction Site (TIS) Model for Single and Double-stranded DNA}

\begin{document}
\begin{abstract}
We develop a robust coarse-grained model for single and double stranded DNA by  representing each nucleotide by three interaction sites (TIS) located at the centers of mass of sugar, phosphate, and base. The resulting TIS model includes base-stacking, hydrogen bond, and electrostatic interactions as well as bond-stretching and bond angle potentials that account for the polymeric nature of DNA. The choices of force constants for stretching and the bending potentials were guided by a Boltzmann inversion procedure using a large representative set of DNA structures extracted from the Protein Data Bank.  Some of the parameters in the stacking interactions were calculated using a learning procedure, which ensured that the experimentally measured melting temperatures of dimers are faithfully reproduced. Without any further adjustments, the calculations based on the TIS model reproduces the experimentally measured salt and sequence dependence of the size of single stranded DNA (ssDNA), as well as the persistence lengths of poly(dA) and poly(dT) chains . Interestingly, upon application of mechanical force the extension of poly(dA) exhibits a plateau, which we trace to the formation of stacked helical domains. In contrast, the force-extension curve (FEC) of poly(dT) is entropic in origin, and could be described by a standard polymer model. We also show that the persistence length of double stranded DNA, formed from two complementary ssDNAs with one hundred and thirty base pairs,  is consistent with the prediction based on the worm-like chain.  The persistence length, which decreases with increasing salt concentration, is in accord with the Odijk-Skolnick-Fixman theory intended for stiff polyelectrolyte chains near the rod limit. The range of applications, which did not require adjusting any parameter after the initial construction based solely on PDB structures and melting profiles of dimers, attests to the transferability and robustness of the TIS model for ssDNA and dsDNA. 

\end{abstract}

\section{Introduction}

DNA, the blueprint of life, accomplishes its functional roles through highly orchestrated motions, spanning a hierarchy of time and length scales. \cite{sanger} Evolution has endowed DNA with high adaptability,  allowing it to undergo conformational changes, in response to cellular cues, without being irreversibly damaged. 
The advances in experimental methodology, in particular, single molecule techniques, have provided critical insight into DNA biophysics, including
various aspects of its structural organization, and sequence-dependent mechanical tensegrity.\cite{seq_mechanics1}  Nonetheless, the physical principles that underlie key attributes of DNA at all length scales, ranging from few hundred base pairs to large scale chromatin organization are not understood.\cite{chromatin_organisation} The growing interest in  DNA nanotechnology,  and the need to formulate design rules for self-assembly, as well as nanofabrication further necessitates  an understanding of DNA thermodynamics, and mechanics. \cite{dna_nanotech2,dna_nanotech1} In all these areas well-designed computational models with sufficient accuracy are needed to provide not only insights into the biophysics of DNA but also for making predictions, especially where experiments cannot fully decipher the sequence-dependent properties of DNA. 

\begin{figure}
\centering
\includegraphics[width=0.65\textwidth]{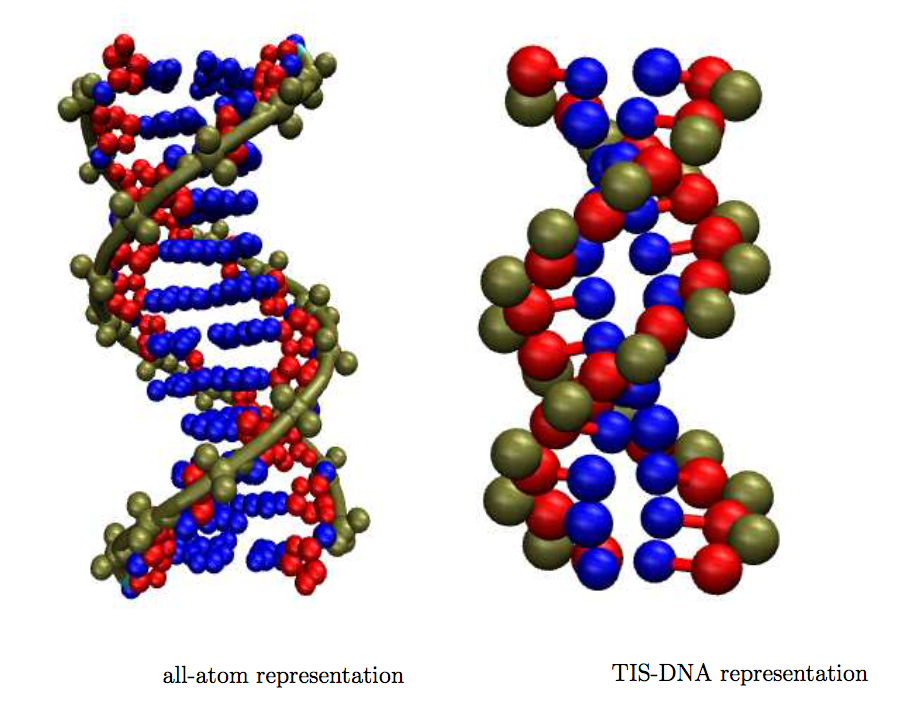}
\caption{The coarse-graining procedure underlying the TIS-DNA model. Each nucleotide is represented by three beads: one for the sugar, base, and the phosphate. These residues are represented using the same color code in the all-atom, and the TIS-DNA representations. As shown above, in the case of a twelve base pair duplex, the number of degrees of freedom reduces from 1458 to 210 upon coarse-graining.}
\end{figure}

A reliable structural model of DNA is required to accurately describe the key features of DNA biophysics at the molecular level. It is always tempting to use an all-atom representation of the DNA molecule, as well as the surrounding solvent, and counterions in order to glean microscopic insights into DNA dynamics.\cite{Beveridge2004,Lavery2010} However, the innumerable degrees of freedom, which are coupled together in a complex fashion, often render it practically impossible to probe DNA dynamics over biologically relevant time scales, and length scales using current computer hardware. More importantly, the current force fields are not accurate enough to obtain results that can be compared to experiments. Instead, it is prudent to use a level of description depending on the length scale of DNA and the accuracy of the measurements. \cite{Hyeon11NatComm}. For example, characterization of the organization of chromosome structure can only be done using copolymer models in which each bead represents $\sim$ 1000 base pairs (bps) whereas assembly of DNA hairpins needs a more refined models. In some cases, such as polymerase-DNA complex, much can be learned using a single bead per base pair representation. \cite{Chen09PNAS,Merenduzzo17SoftMatter}  Inspired by the success of simplified models there have been continued efforts towards the development of coarse-graining (CG) procedures, which reduce the number of degrees of freedom significantly. Despite their simplicity, CG models are often accurate at the molecular, as well as the chemical level, and are built with the aim to embody the underlying physics of nucleic acid mechanics, thermodynamics, and kinetics.\cite{HyeonC.andThirumalai2005,CG_plotkin,oxDNA,depablo_3spn2,Maciejczyk2014,Uusitalo2015,Cho09PNAS,papoian_CG,CG_ngyen}

In a broad sense, DNA coarse-grained models are built either using a ``top-down'' or a ``bottom-up" approach.\cite{CG_review,Drukker2000,Dans2010}  While the former is constructed to reproduce experimental trends and large scale behavior, the  latter exploits systematic coarse-graining to match distributions or forces computed using a more detailed model.  Some coarse-grained models often use a combination of both approaches. \cite{Uusitalo2015}

In this work, we adopt a largely ``top-down" strategy to develop a new coarse-grained model for DNA, in which each nucleotide is represented by three interaction sites (TIS). Several previous studies\cite{HyeonC.andThirumalai2005,Sambriski2009,Denesyuk2013,crowding_TIS} have shown that this choice of resolution is sufficient to describe nucleic acid folding, and mechanical response in the presence of an external force or torque. The TIS-DNA model includes sequence-dependent stacking, hydrogen-bonding, and electrostatic interactions that contribute to the overall stability of DNA structures. The TIS CG model for DNA provides an excellent description of the mechanical properties of both ssDNA and dsDNA, over a wide range of salt concentrations, setting the stage for applications to a wide range of problems involving DNA on not too large a length scale.


\section{Methodology}
\subsection{The Three Interaction Site (TIS) DNA model}


In the TIS model for nucleic acids,  first introduced by Hyeon and Thirumalai,\cite{HyeonC.andThirumalai2005} each nucleotide is represented by three spherical beads (interaction sites), corresponding to the phosphate (P), sugar (S), and the base (B).  The beads are positioned at the the center of mass of the chemical groups.  The energy function describing the interactions between the interaction sites in DNA has the same functional form as the TIS-RNA model, developed by Denesyuk and Thirumalai (DT).\cite{Denesyuk2013, crowding_TIS}  The total energy, $U_{T}$,  for a given conformation of the polynucleotide is expressed as a sum of contributions from six components, denoting the bond ($U_{B}$), angular ($U_{A}$), single-stranded stacking ($U_{S}$), hydrogen-bonding ($U_{HB}$), excluded volume ($U_{EV}$), and electrostatic ($U_{E}$) interactions : 

\begin{equation}
U_{T} = U_{B} + U_{A} + U_{S} + U_{HB} + U_{EV} + U_{E}.
\end{equation}

We use harmonic potentials to describe the bond and angular interactions: 

\begin{equation}
U_{B} = k_{r}(r-r_{0})^2,
\end{equation}
\begin{equation}
U_{A} = k_{\alpha}(\alpha - \alpha_{0})^2,
\end{equation}

\noindent
In equations 2 and 3, $r_{0}$ and $\alpha_{0}$ denote the equilibrium bond lengths, and bond angles respectively, and $k_{r}$, and $k_{\alpha}$ are the corresponding force constants.  

The values $r_{0}$ and $\alpha_{0}$ were obtained by coarse-graining an ideal B-form DNA helix.  
To obtain the initial guesses for $k_{r}$, and $k_{\alpha}$, we carried out Boltzmann inversions\cite{IBI} of the corresponding distributions  obtained from experimental structures.  The statistics were collected from three-dimensional structures of DNA helices deposited in the PDB database.  We exclude all X-ray structures with a resolution lower than 2.5\,\AA, DNA molecules consisting of unnatural bases, and DNA-ligand complexes. The PDB ids of the 284 structures, which met the selection criteria, are available upon request.  Some representative distributions corresponding to the coarse-grained bonds, and angles obtained from the PDB database mining, are shown in Figure 2.

\begin{figure}
\includegraphics[width=0.95\textwidth]{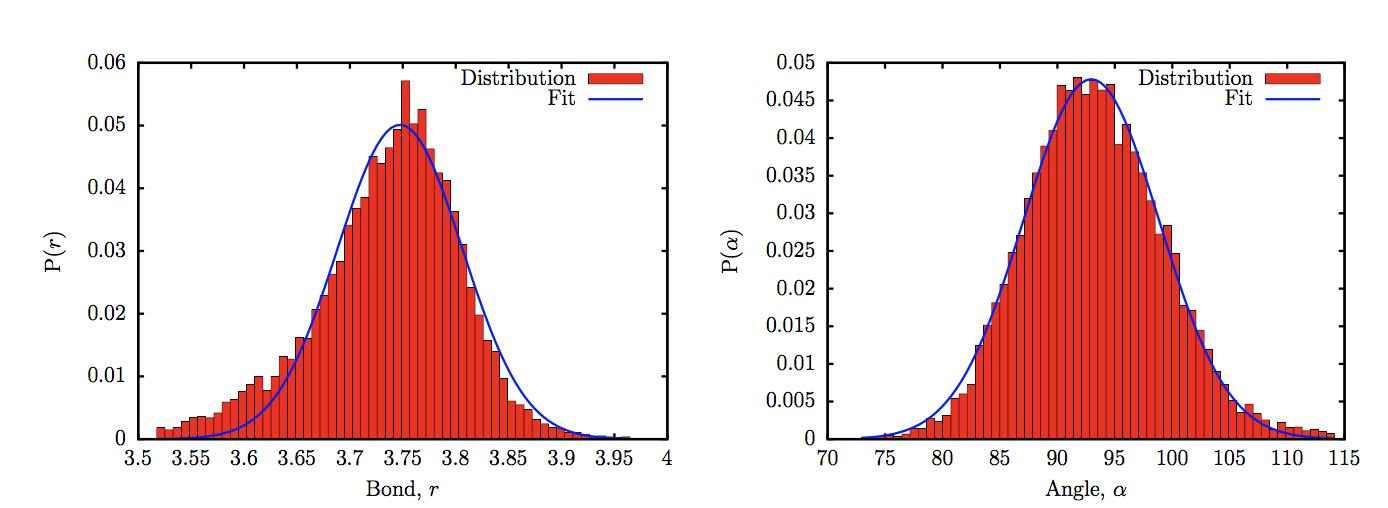}
\caption{Distribution of the SP bond length (left), and SPS bond angles (right) from PDB database mining (red bars). The blue curves are fits to Gaussian functions.}
\end{figure}

\noindent
\textbf{Bond Stretch Potential:} The distribution of the bond lengths can be fit to a Gaussian function:

\begin{equation}
P(r) = \frac{A}{\sqrt{2 \pi \sigma}} e ^{\frac{-\left( r - r_{0} \right)^{2}}{2 {\sigma ^{2}}}} = e^{\frac{-U_{B}}{k_{B}T}},
\end{equation}

\noindent
where the parameter $\sigma$ is obtained from the fit; $k_{B}$ is the Boltzmann factor; and T denotes the absolute temperature.  Taking the logarithm on both sides of (4), and dropping the arbitrary constant, we get: 

\begin{equation}
U_{B} = \frac{k_{B}T}{2 \sigma^{2}}\left( r - r_{0} \right) ^ {2} = k_{r} \left( r - r_{0} \right) ^ {2}.
\end{equation}


\noindent
We estimate $k_{r}$ with $T$ set to 298\,K.  The values of $k_{r}$ for the different bonds are largely insensitive to the choice of $T$, within a broad range. 

\noindent
\textbf{Bond angle potential:} Following earlier work,\cite{Xia2010,Agrawal2014} the distribution of bond angles, $\alpha$,  were weighted by a factor $\sin \alpha $, and renormalized.  The distribution of bond angles is expressed as: 

\begin{equation}
P(\alpha) = f_{n} \frac{p(\alpha)}{\sin \alpha } =  e^{\frac{-U_{A}}{k_{B}T}},
\end{equation}

In (6),  $f_{n}$ is a normalization factor,  while $p \left( \alpha \right)$, and $P \left( \alpha \right)$ denote the unnormalized, and normalized distribution functions, respectively.   The angular potential is obtained using, 

\begin{equation}
U_{A} = \frac{k_{B}T}{2 \sigma^{2}}\left( \alpha - \alpha_{0} \right) ^ {2} = k_{\alpha} \left( \alpha - \alpha_{0} \right) ^ {2}.
\end{equation}

\noindent
\textbf{Excluded Volume:} To account for volume exclusions between the sites, we use the Weeks-Chandler-Andersen (WCA) potential:\cite{WCA} 
 
\begin{equation}
U_{EV} = \epsilon_{0}\left[\left(\frac{D_{0}}{r}\right)^{12} -  2\left(\frac{D_{0}}{r} \right)^{6} + 1\right], r < D_{0}.
\end{equation}

\noindent
The excluded-volume interaction term vanishes if the interacting sites are separated by a distance greater than $D_{0}$, thereby making the WCA potential computationally efficient.  Following DT,\cite{Denesyuk2013} we set $D_{0} = 3.2$\,\AA \,\,and $\epsilon_{0} = 1$\,kcal/mol.   All the interaction sites are assigned the same $D_{0}$ and $\epsilon_{0}$ to keep the parametrization as simple as possible.  As discussed by DT,\cite{Denesyuk2013} this particular choice of $D_{0}$ and $\epsilon_{0}$ somewhat underestimates the distance of closest approach between the interaction sites, with the exception of stacked bases, but has little effect on the folding thermodynamics.

\noindent
\textbf{Stacking Interaction:} Stacking interactions, between two consecutive nucleotides along the DNA chain, is described using the function: 

\begin{equation}
U_{S} =  U_{S}^{0}(1 + k_{l}(l-l_{0})^2 + k_{\phi}(\phi_{1}-\phi_{1}^{0})^2 +  k_{\phi}(\phi_{2}-\phi_{2}^{0})^2)^{-1}
\end{equation}
\noindent
The strength of the stacking interaction  is modulated by deviations from the equilibrium geometry, described by the stacking distance $l_{0}$, and backbone dihedrals $\phi_{1}$, and $\phi_{2}$.  In a previous work, Dima \textit{et al.} showed that an accurate description of stacking in RNA is necessary for fold recognition, and structure prediction.\cite{stacking_Dima} The geometric parameters in terms of which the stacking interactions are represented in the TIS model are described in Figure 3. The equilibrium values for stacking distances and dihedrals are obtained by coarse-graining an ideal B-DNA helix.  We calculated  $k_{l}$ and $k_{\phi}$,  by performing a Boltzmann inversion of the distributions corresponding to the distances between stacked bases ($l$), and backbone dihedrals ($\phi_{1}$, and $\phi_{2}$), computed from the experimental structures. 
\begin{figure}
\includegraphics[width=0.95\textwidth]{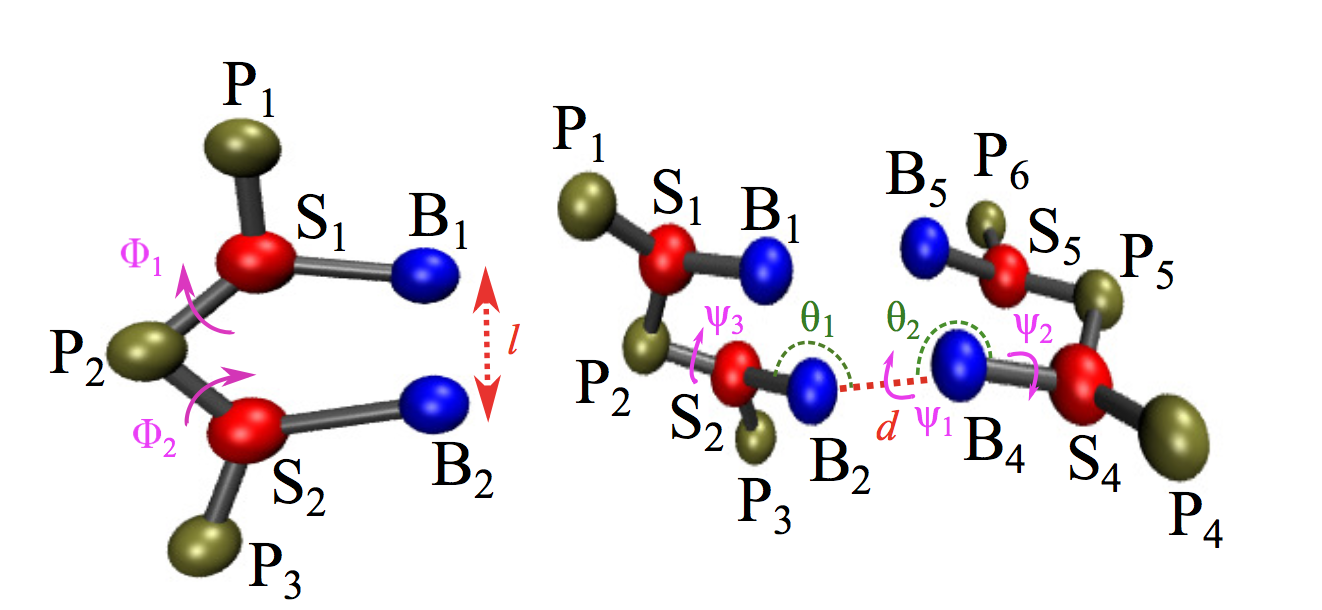}
\caption{Left: Schematic of a coarse-grained dimer based on the TIS model, with $l$, $\phi_{1}$, and $\phi_{2}$ labeled. Right: Illustration of the structural parameters in equation (13). The hydrogen-bonding distance $d$ is between sites $\mathrm{B_{2}}$ and $\mathrm{B_{4}}$; $\theta_{1}$ ($\mathrm{S_{2}}$, $\mathrm{B_{2}}$, $\mathrm{B_{4}}$), and $\theta_{2}$ ($\mathrm{S_{4}}$, $\mathrm{B_{4}}$, $\mathrm{B_{2}}$) are the angles; $\psi_{1}$ ($\mathrm{S_{2}}$, $\mathrm{B_{2}}$, $\mathrm{B_{4}}$, $\mathrm{S_{4}}$), $\psi_{2}$ ($\mathrm{P_{4}}$, $\mathrm{S_{4}}$, $\mathrm{B_{4}}$, $\mathrm{B_{2}}$), and $\psi_{3}$ ($\mathrm{P_{2}}$, $\mathrm{S_{2}}$, $\mathrm{B_{2}}$, $\mathrm{B_{4}}$) are the dihedral angles.}
\end{figure}

In Eq. (9), $U_{S}^{0}$ describes the stacking interaction for a particular dimer, and is calibrated to reproduce the thermodynamics, as described by the nearest-neighbor model.\cite{nearest,Santalucia2004}  In this formalism, the overall stability of DNA duplexes is expressed as a sum over contributions from individual base-pair steps.   Here, we use the unified nearest-neighbor parameters from Santalucia and Hicks (Table 1),\cite{Santalucia2004} which describes the overall stability of duplexes at 1\,M monovalent salt in terms of enthalpic ($\Delta H$), and entropic contributions ($\Delta S$). 

\begin{table}
\centering
\begin{tabular}{l | c | c}
\hline
\hline
& & \\
$\frac{x-y}{w-z}$ & $\Delta H$\, kcal/mol & $\Delta S$\,cal/mol\, K$^{-1}$\\
& & \\
\hline
\hline
$\frac{A-T}{A-T}$ & -7.6 & -21.3\\
& & \\
$\frac{A-T}{T-A}$ & -7.2 & -20.4 \\
& & \\
$\frac{T-A}{A-T}$ & -7.2 & -21.3 \\
& & \\
$\frac{C-G}{A-T}$ & -8.5 & -22.7 \\
& & \\
$\frac{G-C}{T-A}$ & -8.4 & -22.4\\
& & \\
$\frac{C-G}{T-A}$ & -7.8& -21.0 \\
& & \\
$\frac{G-C}{A-T}$ & -8.2& -22.2\\
& & \\
$\frac{C-G}{G-C}$ & -10.6& -27.2\\
& & \\
$\frac{G-C}{C-G}$ & -9.8& -24.4\\
& & \\
$\frac{G-C}{G-C}$ & -8.0& -19.9\\
\hline
\end{tabular}
\caption{\textit{Nearest-neighbor thermodynamic parameters for Watson-Crick base pairs in DNA in 1\,M NaCl and 310\,K.\cite{Santalucia2004} }}
\end{table}

We assume that the $\Delta H$, and $\Delta S$ of each base-pair dimer step can be decoupled into separate contributions arising from single-stranded stacking, and inter-strand hydrogen bonding: 

\begin{equation}
\Delta H  \left( \frac{x - y}{w - z} \right) = \Delta H \left( \frac{x} {w} \right) + \Delta H \left( \frac{z} {y} \right) + 0.5 \Delta H \left( x - y \right) + 0.5 \Delta H \left( w - z\right),
\end{equation}

\begin{equation}
\Delta S \left( \frac{x - y}{w - z} \right) = \Delta S \left( \frac{x} {w} \right) + \Delta S \left( \frac{z} {y} \right).
\end{equation}

\noindent
In Eqs. (10) and (11),  $\Delta H \left( \frac {x} {w} \right)$ and $\Delta S \left( \frac{x} {w} \right)$ denote the enthalpy, and entropy associated with the stacking of $x$ over $w$ in the  5$^{\prime}$ $\rightarrow$ 3$^{\prime}$ direction.  Based on previous experimental data,\cite{kamenetskii} it is reasonable to assume that the contribution from hydrogen-bonding, $\Delta H \left( x - y \right)$, is purely enthalpic in nature.  To obtain the thermodynamic parameters for all the dimers, we need to solve (10) and (11) for $\Delta H \left( \frac{x} {w} \right)$, $\Delta S \left( \frac{x} {w} \right)$, and $\Delta H \left( x - y \right)$.  Since the number of unknowns exceeds the number of equations, we make some additional assumptions based on previous experimental and simulation data. 

\begin{table}
\centering
\begin{tabular}{l | c | c | c}
\hline
\hline
& & \\
$\frac{x}{w}$ & $\Delta H$\, kcal/mol & $\Delta S$\,cal/mol\, K$^{-1}$ & $T_{m}$ (K) \\
& & \\
\hline
\hline
$\frac{A}{A}$ & -3.53 & -10.96 & 322.0\\
& & \\
$\frac{A}{T}; \frac{T}{A}$ & -3.06 & -10.43& 293.0 \\
& & \\
$\frac{A}{G};\frac{G}{A}$ & -3.76 & -11.26& 333.6 \\
& & \\
$\frac{A}{C};\frac{C}{A}$ & -3.06 & -10.43& 293.0 \\
& & \\
$\frac{G}{G}$ & -3.39 & -9.56& 353.9\\
& & \\
$\frac{G}{C};\frac{C}{G}$ & -4.28& -12.90& 331.9 \\
& & \\
$\frac{G}{T};\frac{T}{G}$ & -4.03& -12.13& 332.6\\
& & \\
$\frac{C}{C}$ & -2.98& -10.33& 288.3\\
& & \\
$\frac{C}{T};\frac{T}{C}$ & -2.98& -10.33& 288.3\\
& & \\
$\frac{T}{T}$ & -2.98& -10.33& 288.3\\
\hline
\hline
\end{tabular}
\begin{tabular}{c}
$\Delta H(A-T) = -1.09$\,kcal/mol; $\Delta H(G-C) = -1.64$\,kcal/mol.
\end{tabular}
\caption{\textit{Enthalpies $\Delta H$, entropies $\Delta S$, and melting temperatures $T_{m}$ of single-stranded DNA stacks derived in this work. }}
\end{table}

The thermodynamic parameters corresponding to AT/TA and TA/AT, CA/GT and GT/CA, CT/GA and GA/CT, and CG/GC and GC/CG, as described by equations (10) and (11) can be averaged, as these values are similar within experimental uncertainty.\cite{Santalucia2004}  This enables us to assign $\Delta H \left( \frac{x}{w} \right) = \Delta H \left( \frac{w}{x} \right)$, and $\Delta S \left( \frac{x}{w} \right) = \Delta S \left( \frac{w}{x} \right)$ for all the dimer steps.  Experiments by Olsthoorn \textit{et al.}\cite{OLSTHOORN1981} indicate that the stacking enthalpy of a deoxyadenylate dimer is virtually identical to the ribo analogue.  Hence, we set $\Delta H \left( \frac{A}{A} \right)$ equal to \mbox{-3.53\,kcal/mol,} which is the enthalpy value computed for an adenine-adenine stack in the RNA model.\cite{Denesyuk2013}  The melting temperature ($T_{m}$) of the dimer $\left( \frac{A}{A} \right)$  is estimated to be 322\,K by CD spectroscopy experiments.\cite{OLSTHOORN1981} Furthermore, experiments\cite{sollie} show that the free energy of stacking for a $\left( \frac{T}{T} \right)$ dimer at 298\,K is around 0.1\,kcal/mol. From the above assumptions, and using equations (10) and (11), we can compute $\Delta H \left( \frac{A}{A} \right)$, $\Delta H \left( \frac{T}{T} \right)$, $\Delta S \left(\frac{A}{A}\right)$, $\Delta S \left(\frac{T}{T} \right)$, and $\Delta H\left(A-T\right)$.  Using $\Delta H\left(A-T\right)$, $\Delta H \left(\frac{A}{T}\right)$, and $\Delta S \left(\frac{A}{T}\right)$ are computed from the appropriate thermodynamic equations.  The enthalpies of hydrogen-bonding are related as: $\Delta H \left(G - C\right) = 1.5 \Delta H \left(A - T\right)$.  Once $\Delta H \left( G-C \right)$ is known, $\Delta H \left( \frac{G}{C}\right)$, and $\Delta S \left( \frac{G}{C} \right)$ are computed from equations (10) and (11). 

To evaluate the thermodynamic parameters for the remaining dimers, we need to make additional simplifications.  Experiments by Sollie and Schellman,\cite{sollie} as well as recent simulations\cite{sponer,Jafilan2012,brown_jctc} indicate that $\left( \frac{A}{T} \right)$, and $\left( \frac{A}{C} \right)$ have similar stacking propensities.  Therefore, we can describe them by the same set of thermodynamic parameters: $\Delta H\left(\frac{A}{T}\right) = \Delta H \left(\frac{A}{C}\right)$, and $\Delta S\left(\frac{A}{T}\right) = \Delta S \left(\frac{A}{C}\right)$.  Using the enthalpy and entropy of stacking for the $\left(\frac{A}{C}\right)$ dimer, we estimate the corresponding values for the $\left(\frac{G}{T}\right)$ dimer using the appropriate set of equations from (10) and (11).  We also assume that $\left(\frac{T}{T}\right)$, $\left(\frac{C}{T}\right)$, and $\left(\frac{C}{C}\right)$ dimers can be described by the same set of thermodynamic parameters, as experiments and simulations show that they very similar stacking propensities.\cite{sollie,melvin,Jafilan2012}  This simplification allows us to evaluate $\Delta H \left(\frac{G}{A}\right)$, $\Delta S\left(\frac{G}{A}\right)$, $\Delta H \left(\frac{G}{T}\right)$, $\Delta S\left(\frac{G}{T}\right)$, $\Delta H \left(\frac{G}{G}\right)$, and $\Delta S \left(\frac{G}{G}\right)$.  The results are summarized in Table 2.

\textbf{Using melting temperature of dimers to learn the $U_{S}^{0}$ value:} In order to calibrate the stacking interactions $U_{S}^{0}$, we simulated the stacking of coarse-grained dimers, similar to that shown in Figure 3.  We use the expression for $U_{S}$ in Eq. (9), with $U_{S}^{0} = -h + k_{B}\left( T - T_{m} \right)s$. Here, $h$ and $s$ are adjustable parameters, $T_{m}$ is the melting temperature of a given dimer, as tabulated in Table 2.  In simulations, the free energy of stacking for each dimer can be computed using: 
\begin{equation}
\Delta G = -k_{B}T \ln p + k_{B}T \ln (1-p) + \Delta G_{0},
\end{equation}
where $p$ is the fraction of sampled configurations for which $U_{S} < -k_{B}T$, and $\Delta G_{0}$ is a correction factor that accounts for any difference  in the definition of stacking, and hence $\Delta G$, between experiment and simulation. Since the thermodynamic parameters for the $\left( \frac{T}{T} \right)$ dimer were derived to explicitly match the stacking free energy at 298\,K, we set $\Delta G_{0} = 0$ for this dimer, as well as  for $\left( \frac{C}{C} \right)$,$\left( \frac{C}{T} \right)$, and $\left( \frac{T}{C} \right)$ dimers, all of which are thermodynamically equivalent within our parametrization scheme.   For the rest of the dimers, we choose $\Delta G_{0}$ such that stacking  free energy at 298\,K,  as estimated by osmometry experiments, is reproduced.

\begin{figure}[htbp!]
\includegraphics[width=0.95\textwidth]{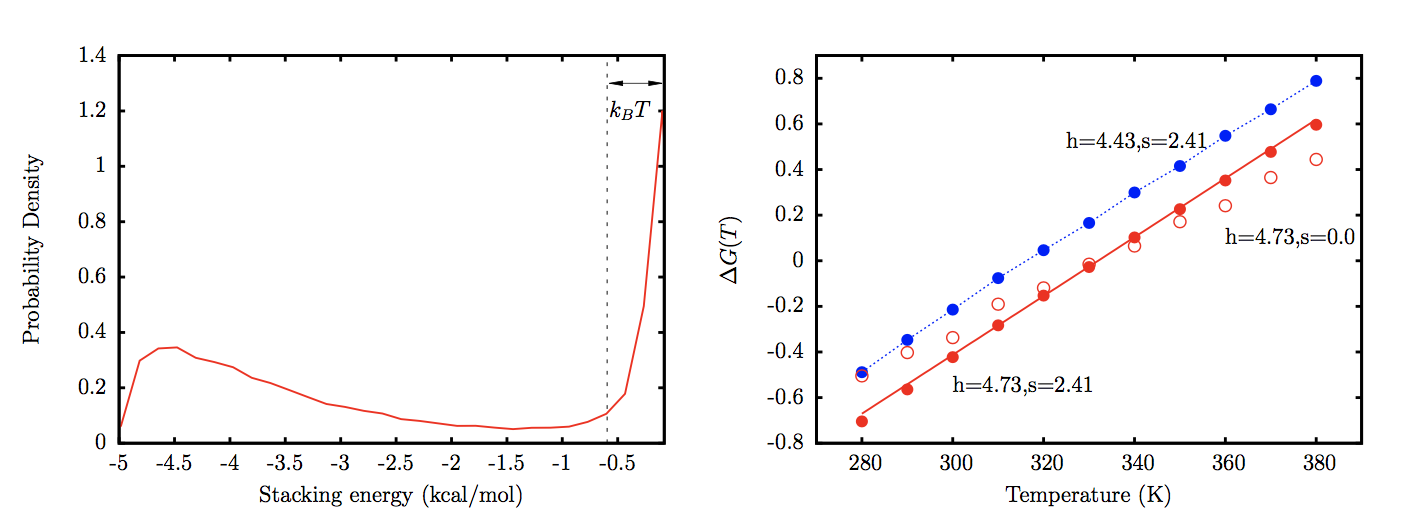}
\caption{Left: A representative distribution of stacking energies, $U_{S}$ from simulations of a dimer represented by the TIS model. All configurations with $U_{S} < -k_{B}T$ are considered as stacked. Right: The free energy of stacking for dimer $\frac{G}{C}$ as a function of temperature. The red solid line corresponds to the experimental line, with free energies given by the parameters in Table 2. The red filled circles represent simulation results, with $h$ and $s$ chosen such that the experimental temperature dependence is reproduced. These values correspond to the case where $\Delta G_{0} = 0$. The red open circles correspond to the case where $T_{m}$ is reproduced, but the entropy is underestimated. The blue circles are simulation data for $h$ and $s$ where the free energy is shifted by a constant factor.}
\end{figure}

\begin{table}[htbp!]
\begin{tabular}{l | c | c | c }
\hline
\hline
$\frac{x}{w}$ & $h$\,kcal/mol & $s$ & $\Delta G_{0}$\\
\hline
& & & \\
$\frac{A}{A}$ &  5.69 (4.67) & 0.94 & 0.60 \\
& & & \\
$\frac{A}{T}$; $\frac{T}{A}$ & 4.95 (4.18); 5.02 (4.28) & 0.87; 0.65 & 0.49  \\
& & & \\
$\frac{A}{C}$; $\frac{C}{A}$ & 4.98 (4.21); 5.03 (4.24) & 0.92; 0.79 & 0.49 \\
& & & \\
$\frac{A}{G}$; $\frac{G}{A}$ & 5.43 (4.83); 5.42 (4.82) & 1.06; 0.98 & 0.39 \\
& & & \\
$\frac{C}{T}$; $\frac{T}{C}$ & 4.13; 4.18 & 0.71; 0.94 & 0.00 \\
& & & \\
$\frac{C}{C}$ &  4.15 & 0.98 & 0.00\\
& & & \\
$\frac{C}{G}$; $\frac{G}{C}$ & 5.22 (4.81); 5.13 (4.73) & 2.33; 2.41 & 0.26   \\
& & & \\
$\frac{G}{T}$; $\frac{T}{G}$ &  5.28 (4.70); 5.43 (4.86) & 1.69; 1.73 & 0.38\\
& & & \\
$\frac{G}{G}$ &  5.66 (5.13) & -0.29 & 0.38\\
& & & \\
$\frac{T}{T}$ &  4.17 & 0.89 & 0.00\\
& & & \\
\hline 
\end{tabular}
\caption{\textit{The parameters $h$ and $s$ are computed from the simulations of coarse-grained dimers. $\Delta G_{0}$ is chosen such that the experimental stacking free energies at 298\,K are reproduced. The values of $h$ before the inclusion of the correction term, $\Delta G_{0}$,  are shown in parentheses.}}
\end{table}


\noindent
Figure 4 shows the results from the simulation of a $\left(\frac{G}{C}\right)$ dimer.  For $\Delta G_{0} = 0$, and $s = 0$,   the melting temperature $T_{m}$ systematically increases with $h$, and is equal to the value in Table 2 for $h = 4.73$.  If $s = 0$, the entropy of stacking, given by the slope of  $\Delta G(T)$ versus $T$ is underestimated compared to the value in Table 2.  To correct for the entropic contribution, we use $U_{S}^{0} = -4.73 + k_{B}(T-331.9)s$, with $s > 0$.  This readjustment does not change $T_{m}$, but allows us to reproduce the entropy, and hence the temperature dependence of $\Delta G(T)$,  in accordance with Table 2.  We find that $s = 2.41$ is an optimal choice for the $\left(\frac{G}{C}\right)$ dimer. The fitting procedure described above is carried out for all the sixteen dimers using the TIS representation.  The final set of parameters is tabulated in Table 3. Some of the dimers, which have equivalent thermodynamic parameters according to our model, have somewhat different $h$ and $s$ values due differences in their equilibrium geometry.

\noindent
\textbf{Hydrogen-bonding interactions:} Hydrogen bonding interactions are only considered between the canonical base pairs (Watson-Crick) in the DNA structure.  In some instances, noncanonical base pairs may play a role in stabilizing the DNA structure.\cite{noncan1,noncan2} Nonetheless, these interactions are excluded from the current model. The CG interaction describing hydrogen-bonding is given by;
\begin{equation}
U_{HB} = \frac{ U_{HB}^{0}}{1 + k_{d}(d-d_{0})^2 + k_{\theta}(\theta_{1} - \theta_{1}^{0})^2 + k_{\theta}(\theta_{2} - \theta_{2}^{0})^2 + k_{\psi}(\psi_{1} - \psi_{1}^{0})^2 + k_{\psi}(\psi_{2} - \psi_{2}^{0})^2 + k_{\psi}(\psi_{3} - \psi_{3}^{0})^2}
\end{equation}

\noindent
where $d$, $\theta_{1}$, $\theta_{2}$, $\psi_{1}$, $\psi_{2}$, and $\psi_{3}$ are described in Figure 3.  The corresponding equilibrium values are obtained from the coarse-grained structure of an ideal B-DNA helix.  The coefficients $k_{l}$, $k_{\theta}$, and $k_{\phi}$ were determined in a fashion similar to the other harmonic constants, using a Boltzmann inversion of the statistics accumulated from experimental structures.  The parameter $U_{HB}^{0}$ controls the strength of the hydrogen-bonding. Similar to base-stacking, hydrogen-bonding is sensitive to deviations from the equilibrium geometry. 

Equation 13 denotes the $U_{HB}$ for a single hydrogen-bond, and is multiplied by a factor of 2 or 3 depending on the type of base pair (A-T or G-C) connecting the coarse-grained sites. 


\noindent
\textbf{Electrostatic interactions: } The electrostatic interactions are computed using the Debye-H{\"u}ckel approximation, in conjunction with the concept of Oosawa-Manning counterion condensation.  The electrostatic free energy is given by:\cite{honig} 

\begin{equation}
U_{E} = \frac{Q^{2}e^{2}}{2\epsilon}\sum_{i,j}\frac{exp(-\vert r_{i} - r_{j} \vert/\lambda_{D})}{\vert r_{i} - r_{j} \vert}
\end{equation}

\noindent
where $\vert r_{i} - r_{j}\vert$ is the distance between two phosphates $i$, and $j$, $\epsilon$ is the dielectric constant of water, and $\lambda_{D}$ is the 
Debye-screening length. The Debye length $\lambda_{D}$ is related to the ionic strength of the solution, and is given by 

\begin{equation}
\lambda_{D} ^{-2} = \frac{4\pi}{\epsilon k_{B}T}\sum_{n}q_{n}^{2}\rho_{n}
\end{equation}

In Eq. (15), $q_{n}$ is the charge for an ion of type $n$, and $\rho_{n}$ is the number density of the ion in solution. 

The magnitude of phosphate charge, $Q$, is determined using the Oosawa-Manning  theory.\cite{manning} The bare charge on the phosphate is renormalized due to propensity of ions to condense around the highly charged polyanion.  The Oosawa-Manning theory predicts that the renormalized charge on the phosphate is

\begin{equation}
Q = Q^{\prime}(T) = \frac{b}{l_{B}(T)}, \quad\text{}\quad  l_{B}(T) = \frac{e^{2}}{\epsilon k_{b} T}.
\end{equation}
\noindent
where $b$ is the length per unit charge, and $l_{B}$ is the Bjerrum length. 
The length per unit charge for DNA, as estimated by Olson and coworkers,\cite{olson_length} is approximately 4.4\,\AA, which leads to a reduced charge of $-$0.6 for the phosphates at 298\,K.  As the dielectric constant is also a function of temperature, the temperature dependence of $Q$ is nonlinear\cite{dielectric_book} with

\begin{equation}
\epsilon(T) = 87.740 - 0.4008\,T + 9.398 \times 10^{-4}\, T^{2} - 1.410 \times 10^{-6}\, T^{3}.
\end{equation}
\noindent
In Eq. 17, $T$ is the temperature in Celsius.  Following DT, the charges are placed on the center of mass of the phosphate beads,\cite{Denesyuk2013} which is somewhat comparable to atomistic representations where the charges are localized on the two oxygen atoms of the phosphate group. 

\subsection{Calculation of persistence length}


The persistence length, a measure of stiffness of DNA, is calculated using the decay of the autocorrelation of tangent vectors $t \hat(s)$ along the backbone. For a worm-like chain (WLC), such as DNA,\cite{doi_edwards} 

\begin{equation}
\langle t\hat(s) t\hat(0)\rangle = \langle cos(\theta(s)) \rangle = exp\left(\frac{-s}{l_{p}}\right).
\end{equation}

\noindent
In equation (18), $\langle ... \rangle$ denotes an average, $s$ denotes position along the DNA strand, and $l_{p}$ is the persistence length.  For ssDNA,  the tangent $t\hat(s)$ was calculated by taking the distance vector from the sugar bead  on nucleotide $i$ to the sugar bead on nucleotide $i+1$, and normalizing it to unity.  For dsDNA,  the tangent vector $t\hat(s)$ was calculated by taking the distance vector from the midpoint of the bases involved in hydrogen bonding at position $i$ along the chain, to the midpoint of the bases at position $i+5$, and normalizing it to unity.  We found that the values of the correlations were quite insensitive to our particular definition of tangent vectors. 

Although the relationship described by Eq. (18) is quite robust for dsDNA, it breaks down when the decay of the autocorrelation function  becomes non-exponential. This situation typically arises for charged flexible polymers, such as ssDNA.\cite{doi_edwards}  Hence, we use the following relationship to estimate $l_{p}$ for ssDNA, following Doi and Edwards.\cite{doi_edwards} 

\begin{equation}
l_{p} = \frac{\langle R^{2} \rangle}{2l},
\end{equation}
where $R$ is the end-end distance, and $l$ is the contour length of the ssDNA chain.






The persistence length of a polyelectrolyte chain, such as DNA, exhibits a strong dependence on the ionic strength of the solution.\cite{persistence_length_expt,Ha1995}  It is known that polyelectrolytes (PEs), such as DNA chain, become more flexible with an increase in ionic strength due to a more effective screening of the phosphate-phosphate charge repulsion resulting from counterion condensation. Nonetheless, the interplay between the DNA and PE effects in determining  the overall chain stiffness is not known. For a stiff PE near the rod limit, the Odijk-Skolnick-Fixman (OSF) theory\cite{osf_1,osf_2} provides a very good description of the electrostatic contribution to persistence length:\cite{esp_Ha, esp_Netz}

\begin{equation}
l_{p} = l_{p0} + l_{OSF}  = l_{p0} + \frac{\lambda_{D}^{2}}{4 \l_{B}}
\end{equation}
where $l_{p0}$ is the bare persistence length, which depends on the intrinsic geometric properties of the chain, $\lambda_{D}$, and $\l_{B}$ denote the Debye length, and Bjerrum length, respectively. In the OSF theory, whose validity extends to flexible polyelectrolytes as well, \mbox{it is assumed that $l_{p0} > l_{OSF}$.}

\subsection{Langevin Dynamics Simulations} The equations of motion of each bead is described by Langevin dynamics, which for bead $i$ can be expressed as a stochastic differential equation:  $m_{i}\bm{\ddot r}_{i} = -\gamma_{i} \bm{\dot r}_{i} + \bm{F}_{i} + \bm{g}_{i}$, where $m_{i}$ is the mass of the bead, $\gamma_{i}$ is the drag coefficient, $\bm{F}_{i}$ denotes the conservative force acting on bead $i$ due to  interactions with the other beads,  and $\bm{g}_{i}$ is a Gaussian random force.  The random force satisfies $\langle \bm{f}_{i}(t) \bm{f}_{j}(t^{\prime})\rangle = 6k_{B}T \gamma_{i} \delta_{ij} \delta(t-t^{\prime})$.  A variant of the velocity-Verlet version of the algorithm for Langevin dynamics,\cite{honeycutt_dt} with a time step of 2.5\,fs, was used to integrate the equations of motion. For the mechanical pulling simulations, we use a time step of 1.25\,fs to maintain the stability of the system. The drag coefficient corresponding to each bead, $\gamma_{i}$,  is calculated using the Stokes' formula, $\gamma_{i} = 6\pi \eta R_{i}$, where $\eta$ is the viscosity of the surrounding environment,  and $R_{i}$ is the Stokes' radius.  We used a value of $10^{-5}$ Pa.s for $\eta$, which is around 1\% of the viscosity of water. This choice does not affect the thermodynamic properties, but is critical for an efficient exploration of the conformational space. \cite{Denesyuk2013,honeycutt_dt} The values for $R_{i}$ are 2\,\AA\, for the phosphate beads,  2.9\,\AA\, for sugar beads, 3\,\AA\, for guanine beads, 2.8\,\AA\, for adenine beads, and 2.7\,\AA\, for cytosine and thymine beads.  Each simulation was carried out for at least $4 \times 10^{8}$ time steps.  To obtain meaningful statistics for any given observable, we carried out at least five simulations for each data point, with different initial conditions. 

\subsection{Parametrization of the DNA model}

\noindent
\textbf{Bonded Interactions:} The range of harmonic constants ($k_{r}$, $k_{\alpha}$) was obtained using equations (5) and (7). To parametrize the coarse-grained DNA model, in terms of mechanical properties, we chose a heterogeneous single-stranded DNA sequence (CATCCTCGACAATCGGAACCAGGAAGCGCCCCGCAACTCTGCCGCGATCGGTGTTCGCCT) with 60 nucleotides. The objective was to optimize the angular bending constants ($k_{\alpha}$), in particular, such that the persistent lengths, computed at different monovalent salt concentrations, fell within the experimental range.\cite{Murphy2004,Kuznetsov2001,Doose2007,Chen2012} During the parametrization process, we switched off the stacking interactions. Besides eliminating the complexity arising due to base-stacking, this choice enabled us to compare our simulated results with persistence length estimates for unstructured ssDNA available from recent experiments.\cite{Chen2012} The choice of bond-stretching constants, $k_{r}$, had practically no effect on the persistence length estimates.  Once an optimal set of harmonic constants were identified,  the stacking interactions were parametrized using the procedure described earlier.  The final set of parameters, employed in our coarse-grained model is tabulated in Table 4. 

\begin{table}
\begin{tabular}{l | l | l}
\hline
Bond Type & force constant,$k_{r}$(kcal/mol/\AA) & equilibrium value,$r_{0}$ (\AA) \\
\hline
SP & 62.59 & 3.75\\
PS & 17.63 & 3.74 \\
SA & 44.31 & 4.85\\
SG & 48.98 & 4.96\\
SC & 43.25 & 4.30\\
ST & 46.56 & 4.40\\
\end{tabular}


\begin{tabular}{l | l | l}
\hline
Angle Type & force constant,$k_{\alpha}$ (kcal/mol/rad) & equilibrium value,$\alpha_{0}$ (degrees)\\
\hline
PSP & 25.67 & 123.30\\
SPS & 67.50 & 94.60\\
PSA & 29.53 & 107.38\\
PST & 39.56 & 97.18\\
PSG & 26.28 & 111.01\\
PSC & 35.02 & 101.49\\
ASP & 67.32 & 118.94\\
TSP & 93.99 & 123.59\\
GSP & 62.94 & 116.90\\
CSP & 77.78 & 121.43\\
\end{tabular}
\newpage
\begin{tabular} {l | l}
\hline
$k_{l}$ (\AA$^{-1}$) & 1.45 \\
$k_{\phi}$ (radians$^{-1}$) & 3.00\\
\hline
$k_{d}$ (\AA$^{-1}$) & 4.00 \\
$k_{\theta}$ (radians$^{-1}$) & 1.50\\
$k_{\psi}$ (radians$^{-1}$) & 0.15\\
\hline
\end{tabular}
\caption{\textit{The optimal values for the various harmonic force constants in the DNA model. The $k_{r}$, $k_{l}$, $k_{\phi}$, $k_{d}$, $k_{\theta}$, $k_{\psi}$ values correspond to those which are obtained directly from the Boltzmann inversion of distributions obtained from database mining.  No further optimization of these parameters were necessary. The angle-specific $k_{\alpha}$ values correspond to those that reproduce the experimental persistence lengths of ssDNA, and were obtained after fine-tuning of the initial parameters obtained from Boltzmann inversion.}}
\end{table}

\noindent

\textbf{Calibration of hydrogen-bonding interaction:} After the optimization procedure, the only free parameter in the model is $U_{HB}^{0}$.  We chose its value to reproduce  the experimental melting curve\cite{ansari_hairpin} at 0.25\,M for  a DNA hairpin with the sequence (shown in the inset of Fig. 5).  The relatively small size of the hairpin, heterogeneity of the stem composition, as well as extensive thermodynamic data available for this hairpin, \cite{ansari_hairpin, ansari_hairpin2} make it an ideal candidate for our calibration procedure.  

\begin{figure}
\includegraphics[width=0.95\textwidth]{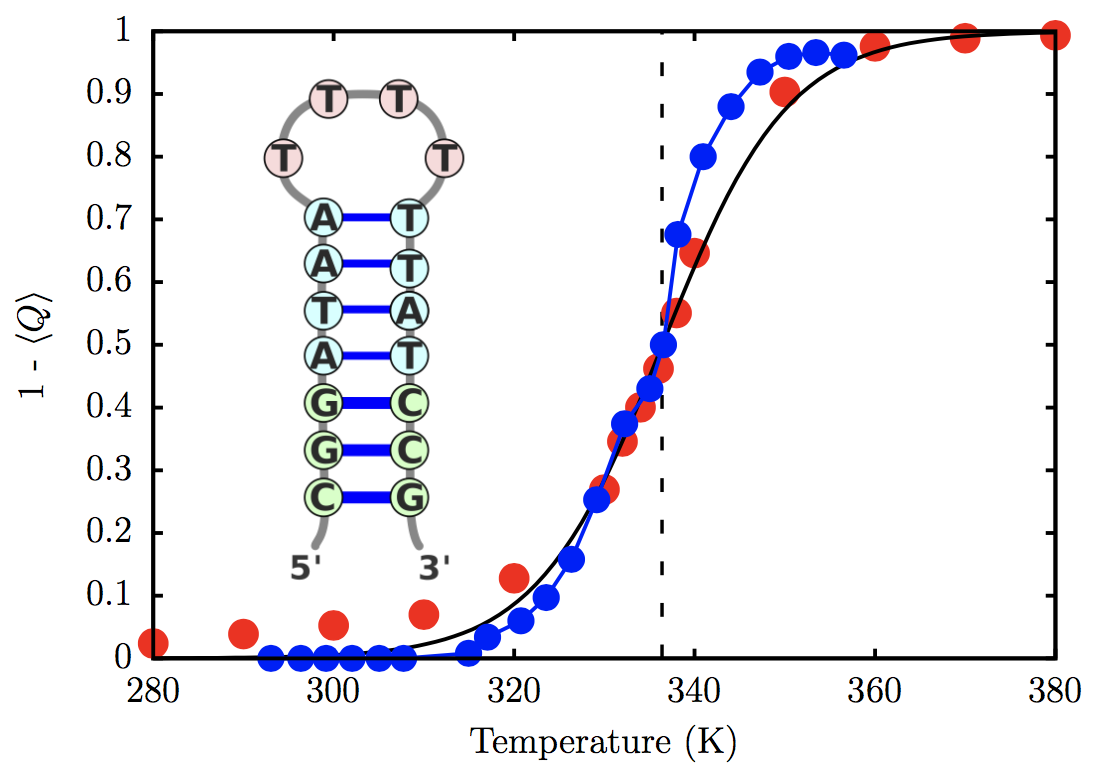}
\caption{Variation of the fraction of broken base pairs, $1 - \langle Q \rangle$, with temperature. A quantitative agreement with experimental melting profile is obtained for $U_{HB}^{0} = -1.92$\,kcal/mol. The simulation results are given in red circles. The experimental estimates \cite{ansari_hairpin} correspond to blue circles. The solid line is a sigmoidal fit to the simulation data, which gives $T_{m}$ = 336.4\,K. The dashed vertical line indicates the experimental melting temperature.\cite{ansari_hairpin}}
\end{figure}

In the experiment, the increase of the relative absorbance with temperature corresponds to both unstacking of bases, as well as breakage of hydrogen bonds. For the  hairpin sequence considered here, the former effect is minimized due to the weak stacking interactions between the thymine bases.\cite{ansari_hairpin} The loss of hydrogen bonding occurs in a largely cooperative fashion,  and at the melting temperature approximately half of the base pairs are broken.  In our model, we consider a hydrogen bond to be formed between the coarse-grained sites if $U_{HB} < -nk_{B}T$, where $n$ = 2 for a A-T base pair, and 3 for a G-C base pair.  Using this definition, we can compute $\langle Q \rangle$, the fraction of native contacts as a function of temperature.  Assuming that $\langle Q \rangle$ is an appropriate order parameter for describing DNA hairpin thermodynamics,  we can determine the melting temperature, $T_{m}$ from the following sigmoidal fit: 
\begin{equation}
1-\langle Q \rangle = \frac{1}{1+e^{-\sigma(T-T_{m})}}.
\end{equation}
\noindent
In the above equation, $\sigma$ is the width of the melting transition.  We find that for $U_{HB}^{0}$ = -1.92\,kcal/mol, the TIS-DNA model reproduces the experimental curve \mbox{(Figure 5)}.  Using equation (21), we estimate $T_{m} = 336.4$\,K, which exactly corresponds to the experimental estimate.\cite{ansari_hairpin}  The width of the transition is slightly overestimated compared to experiment.  This discrepancy likely caused by the neglect of non-native base pairs, as well as anisotropic interactions in our model. Similar deviations were also observed in previous studies by Dorfman and coworkers.\cite{dorfman1,dorfman2}

 
\section{Results and Discussion}

It should be pointed out that the parameters in our TIS-DNA model were determined using statistics generated from PDB structures, and thermodynamic properties of dimers. This is the same learning procedure used by DT to probe the thermodynamic properties of RNA folding.\cite{Denesyuk2013}  The TIS-DNA force field was not calibrated using experimental data in the applications described below. Therefore, the results are genuine predictions of the model. The success, as assessed by comparison to experiments, provides the much needed validation.


 

\subsection{Description of single-stranded DNA}

In the following sections, we describe the applications of the TIS-DNA model to obtain the sequence and salt-dependent mechanical, as well as thermodynamic properties of single-stranded DNA.  We compare the predictions of our model to available experimental data, or well-established theoretical results. 

\subsection{Radius of gyration: salt-dependent scaling behavior}


The dependence of radius of gyration ($\mathrm{R_{g}}$) on the length of a flexible polymer chain is often described by an universal Flory scaling law,\cite{de_gennes} $R_{g} = A_{0} N^{\nu}$,  where $N$ is the number of segments, and $\nu$ is the Flory exponent.   An ideal chain with $\nu = 0.5$, and a rigid rod with $\nu = 1$ denote two limiting cases.  
For a random coil, with excluded volume, the scaling exponent is predicted to be $\sim$\,0.6, based on renormalization group based approaches. \cite{doi_edwards,renormal}

\begin{figure}[!htbp]
\includegraphics[width=0.95\textwidth]{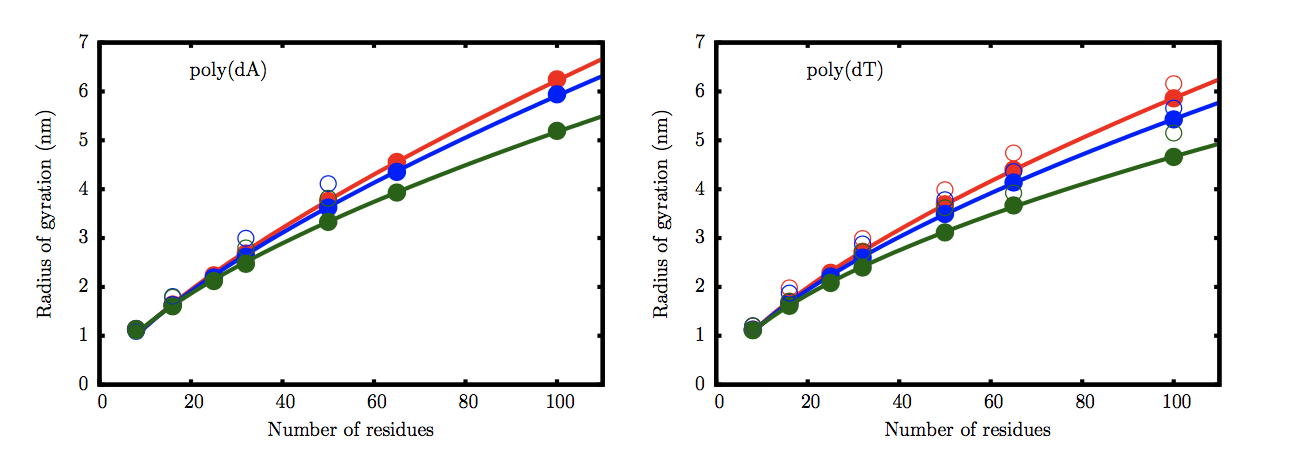}
\caption{Dependence of the radius of gyration $R_{g}$ on the number of nucleotides for a dA (left), and dT oligomer (right). The filled circles  with errorbars are the simulation results at salt concentrations of 0.125\,M (red), 0.225\,M (blue), and 1.025\,M (green). The open circles are experimental data from Sim \textit{et al.}\cite{Sim2012}. The solid lines are fits to the scaling law with $R_{g} = A_{0}N^{\nu}$. }
\end{figure}




In Figure 6, we illustrate the dependence of $R_{g}$ on the chain length, for single-stranded dA and dT  sequences, as described by the TIS-DNA model.  Data are shown for three salt concentrations. As the salt concentration is increased, the power law dependence becomes weaker for both the ssDNA sequences.  This trend is typical of charged polymers, where an increase in chain collapsibility at high ionic strengths results from a more effective screening of the backbone charges. Overall, the predicted values are in good agreement with the estimates from small angle X-ray scattering (SAXS) experiments.\cite{Sim2012}, in contrast to most currently available DNA models,\cite{Uusitalo2015,ssDNA_aa} which lead to over compaction of ssDNA.

It is clear that the TIS-DNA model provides an excellent description of the sequence-dependent variation of the scaling exponents,\cite{Sim2012} $\nu$, with salt concentration (Figure 7).  For the dT sequence, a fit of the simulation data to the scaling law yields \mbox{$\nu \sim 0.67$}, at 0.125\,M. The effective scaling exponent decreases in an exponential fashion with increasing salt concentration, and falls below the random coil limit ($\nu = 0.588$) at around 1\,M.  Therefore, our model predicts that in the moderate to high salt regime, poly(dT) behaves as a random coil, which is in accord with recent experimental findings.\cite{Sim2012,pollack_saxs}

\begin{figure}[!htbp]
\includegraphics[width=0.85\textwidth]{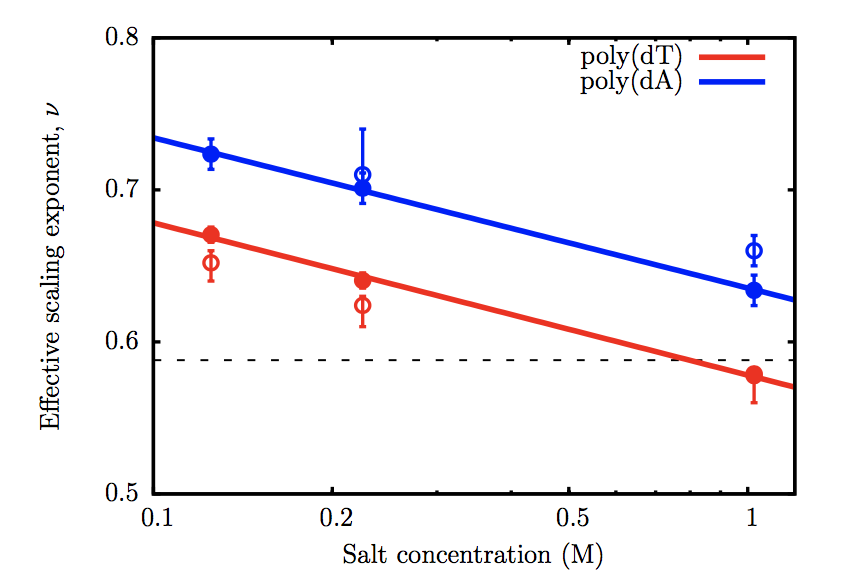}
\caption{The variation of the effective scaling exponent, $\nu$, with salt concentration. The filled circles correspond to $\nu$ obtained from power-law fits of $R_{g} = A_{0}N^{\nu}$. The scaling exponents for both poly(dA) and poly(dT) decrease exponentially with salt concentration.  The open circles denote $\nu$ at different salt concentrations, reported by Sim \textit{et al.}\cite{Sim2012} The dashed line is the scaling exponent ($\nu=0.588$) for a random coil, with excluded volume interactions. }
\end{figure}


The effective scaling exponents for the poly(dA) sequence are consistently higher than for poly(dT) at all salt concentrations. Interestingly, even at 1\,M, the poly(dA) chain does not display random coil-like behavior, unlike poly(dT).  Within the Debye-H{\"u}ckel approximation employed in our model, the two ssDNA sequences have the same charge densities for a given chain length, and therefore electrostatics is unlikely to result in such disparate behavior. Previous work,\cite{base_stacking_dA,fex1} suggest that the origin of this contrasting flexibilities lies in the  chemical difference between adenine and thymine: while the former exhibits significant stacking propensity, base-stacking is disfavored in the latter.  


The preexponential factors, $A_{0}$, obtained from the power-law fits, lie within the experimental range.\cite{Sim2012,Kohn2004,Wilkins99} In the salt concentration range from 0.1 to 1\,M, $A_{0}$ varies from 0.26 to 0.32\,nm for poly(dT), and from 0.22 to 0.27\,nm for poly(dA) sequences. The specific values usually depend on the chemical, and geometric details of the monomer. For the poly(dA) sequence,  the systematically lower $A_{0}$ values at all salt concentrations imply that the effective monomer-monomer bond length (in this particular case,  the distance between two consecutive nulceobases) is shortened as a consequence of base stacking.  On the other hand, the preponderance of unstacked monomers in poly(dT) chains, results in higher $A_{0}$ values, and consequently leads to a stronger dependence of $R_{g}$ on salt concentration.

The distinct stacking property exhibited by adenine, and thymine, is an emergent feature of the TIS-DNA model, as it accounts for base-step dependent stacking thermodynamics.  The presence of base-stacking interactions reduces the collapsibility of the poly(dA) chain, and therefore the $R_{g}$ dependence on chain length follows a stronger power-law compared to poly(dT).  In subsequent sections, we discuss in more detail how the presence of persistent helical stacks along the poly(dA) chain could affect its mechanical tensegrity, and lead to signatures (``plateau") in the force-extension profile.

\subsection{Sequence dependent stiffness}



Despite the ongoing efforts,  some ambiguity exists regarding the persistence length ($l_{p}$) of ssDNA in solution.  The reported values of $l_{p}$ span a wide range, from 1.0 to 6.0\,nm, and is often sensitive to the experimental setup.\cite{Murphy2004,Chen2012,Kuznetsov2001,tinland_ssDNA}  To further validate the robustness of the TIS-DNA model in describing ssDNA, we compute $l_{p}$ using equation (19) for a homogeneous poly(dT), and a poly(dA) sequence, each of which is 40 nucleotide long.  

As outlined in the Methodology section, exponential fits to the decay of tangent correlations provide another means to estimate persistence lengths. Nonetheless, we find that this method of computing $l_{p}$ breaks down for ssDNA, as noted in previous studies. \cite{biexp_fit,Toan2012} The polyelectrolyte nature of DNA is primarily responsible for this deviation. In a previous work, Toan and Thirumalai\cite{Toan2012} showed that tangent correlations decay as a power law, rather than an exponential fashion, over length scales shorter than the characteristic Debye length.

\begin{figure}
\includegraphics[width=0.85\textwidth]{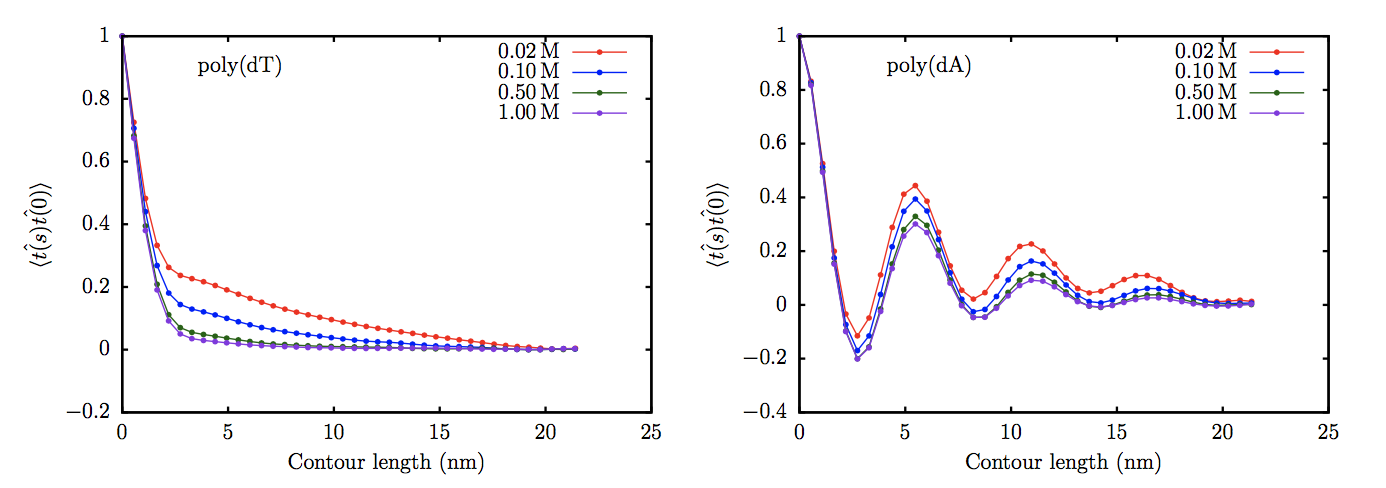}
\caption{The decay of tangent correlations with the contour length in ssDNA for the dT40 sequence (left), and dA40 sequence (right). The correlation function corresponding to dA40 shows oscillatory behavior indicating the presence of helical structure within the chain.}
\end{figure}

The autocorrelation function of the tangent vectors computed at different salt concentrations (Figure 8), show dramatic differences in the equilibrium conformations adopted by the two ssDNA sequences. For poly(dA), the decay of the tangent correlations exhibits oscillatory behavior, which is characteristic of helical structure formed by base-stacking interactions within the chain.\cite{pollack_saxs} No such signatures are observed for the poly(dT) sequence, within a range of salt concentration, suggesting that the corresponding equilibrium ensemble is largely unstructured.  The tangent correlations decay in a non-exponential fashion, particularly at low salt concentrations ($\sim$ 0.02 to 0.10\,M), where electrostatic effects are dominant.  From Figure 9, it is evident that the decay of the tangent correlations becomes exponential over long length scales, but exhibits substantial curvature at short distances. Specifically, the power law behavior postulated by Toan and Thirumalai\cite{Toan2012} becomes apparent when the data is plotted on the logarithmic scale.

\begin{figure}
\includegraphics[width=0.95\textwidth]{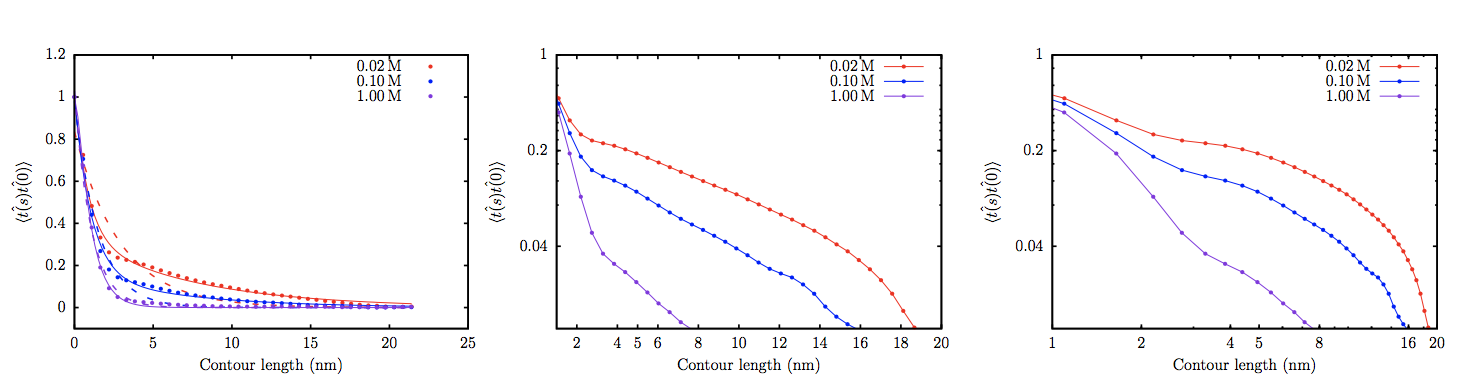}
\caption{Left: The decay of tangent correlations with contour length for the poly(dT), at different ionic strengths. The dashed curves denote fits to a single exponential. The solid lines denote fits to a double exponential. Middle: A semi-log plot of the data, showing the exponential decay at large length scales. Right: Log-log plot of the tangent correlations, showing evidence of power law behavior.}
\end{figure}


\begin{figure}[!htbp]
\includegraphics[width=0.95\textwidth]{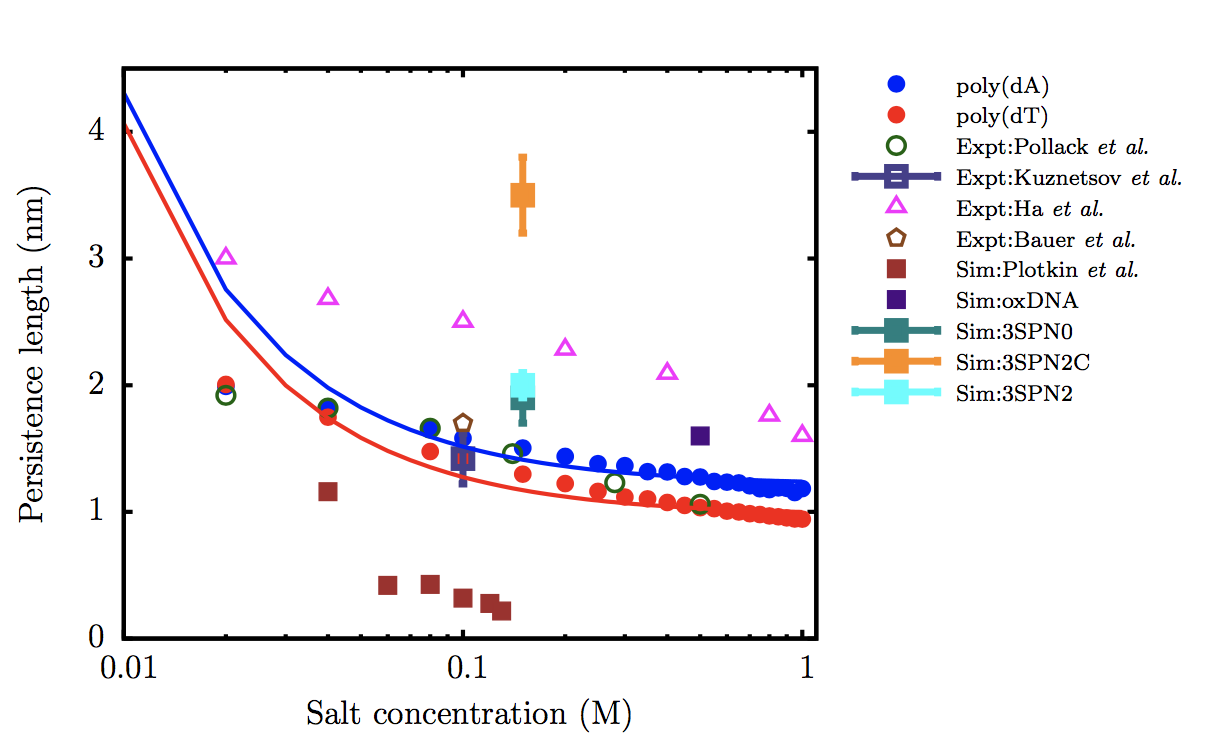}
\caption{The variation of the persistence length ($l_p$) with salt concentration for the dT40 (red) and dA40 (blue) sequences  computed from the simulation. The solid curves are fits of the simulation data to the OSF theory (Eq. (20). The $l_{p}$ values reported by Pollack and coworkers,\cite{Chen2012} using  a combination of SAXS and smFRET experiments, are shown as green open circles. The triangles denote the experimental data of Ha and coworkers.\cite{Murphy2004}. The open square with error bar represents the experimental data of Kuznetsov \textit{et al.},\cite{Kuznetsov2001} for a dT hairpin at 0.1\,M. The brown polygon denotes the $l_{p}$ value reported by Bauer and coworkers\cite{Doose2007} for a dT100 sequence at 0.1\,M using FCS experiments. The filled squares denote the persistence lengths estimated by using coarse-grained models with a resolution similar to ours:  data in brown are from Plotkin and coworkers\cite{CG_plotkin}, oxDNA (purple),\cite{oxDNA} 3SPN0 (teal),\cite{depablo_3spn0} 3SPN.2 (cyan),\cite{depablo_3spn2} and 3SPN2C(orange).\cite{depablo_3spn2c} We do not include the data for 3SPN1\cite{Sambriski2009} as it lies outside the experimental range.} 
\end{figure}


 
As shown in Figure 10, the persistence length, $l_{p}$, of ssDNA predicted by the TIS-DNA model falls within the range of experimentally reported values, over the entire range of salt concentration.  In particular, the agreement between our estimates for poly(dT) \mbox{(red circles)}, and those reported in a recent study by Pollack and coworkers\cite{Chen2012} using a combination of SAXS and smFRET experiments (green open circles), is remarkable. At a salt concentration of 0.1\,M, our model predicts $l_{p}$ to be 1.45\,nm, which lies within the error bars of the values determined by Kuznetsov \textit{et al.} (1.42\,nm),\cite{Kuznetsov2001} and Bauer and coworkers (1.7\,nm),\cite{Doose2007} respectively.  Our results for the dT40 sequence deviate from the experimental values of Ha and coworkers,\cite{Murphy2004} particularly at low salt concentrations.  This discrepancy was also noted in the experimental study of Pollack and coworkers,\cite{Chen2012} who ascribed the reason to the disparate boundary conditions.  In their experiment, Ha and coworkers\cite{Murphy2004} measured $l_{p}$ for a  ssDNA construct, which was attached to the end of a long DNA duplex.  This tethering  impedes the motion of the ssDNA segment, and likely alters the values of $l_p$. In contrast, in the experiment of Pollack and coworkers,\cite{Chen2012} the ssDNA molecules diffuses freely, and this setup is commensurate with the boundary conditions employed in our simulations.  Besides agreement with the experimental data, the variation of $l_{p}$ with salt concentration is in accord with the OSF theory. A fit to the OSF equation yields a bare persistence length $l_{p0}$ of 0.98\,nm for poly(dT), which is within the range (0.6 to 1.3\,nm) of values reported by different experiments.\cite{bare_ssDNA,bare_ssDNA2,tinland_ssDNA,Chen2012} 
  
The poly(dA) sequence is stiffer than poly(dT),  particularly at high salt concentration, where the electrostatics are effectively screened, and the intrinsic geometric nature of the chain manifests itself.  A fit to the OSF theory yields a bare persistence length of 1.22\,nm.  The organization of neighboring bases into stacked helices in poly(dA) endows the ssDNA chain with additional bending rigidity, leading to systematically larger $l_{p}$ values.  Our prediction is also consistent with the experiment of Goddard \textit{et al.},\cite{Goddard} who found that hairpin formation in poly(dA) involves a larger enthalpic cost, compared to poly(dT). The contrasting persistence lengths estimated for poly(dT) and poly(dA)  further explain why these sequences exhibit entirely different salt-dependent collapse, as explained in the earlier section.  

It should not go unstated that the results of the TIS-DNA model  represents a significant improvement over other currently available coarse-grained DNA models of similar resolution in describing the flexibility of ssDNA.  As shown in Figure 10, the model of Plotkin and coworkers\cite{CG_plotkin} severely underestimates the $l_{p}$ of ssDNA over the entire range of salt concentration. On the other hand the latest version of the 3SPN model\cite{depablo_3spn2c}, which was optimized for DNA duplex structures, predicts ssDNA to be too stiff.  Although the older versions of the 3SPN model,\cite{depablo_3spn0,depablo_3spn2} and oxDNA\cite{oxDNA} estimate $l_{p}$ values that fall within the experimental range, the bare persistence lengths, $l_{p0}$, is overestimated. These discrepancies would produce incorrect force-extension behavior, and severely limits the use of such models in applications where a correct description of ssDNA flexibility is important.

\subsection{Sequence-dependent force-extension profiles of ssDNA}

Single-molecule pulling experiments provide a viable route towards determining the mechanical properties, as well as the thermodynamics of base-stacking in DNA.  The applied force, in conjunction with electrostatic repulsion between the backbone phosphates destabilizes base-stacking, and the subtle interplay between competing interactions leads to a variety of elastic regimes.  Several studies report that the tensile response of ssDNA is dictated by sequence composition, as well as ionic strength.\cite{fex1,fex2,fex3} Recently it was shown in designed homogeneous sequences, such as poly(dA), which exhibit substantial base-stacking, the elastic response deviates substantially from the predictions of the standard polymer models.\cite{marko_siggia,halperin}.  Specifically, a plateau in the low-force regime of the force-extension profile is thought to be the `mechanical footprint' of base-stacking. \cite{fex1,fex2}


\begin{figure}[!htbp]
\includegraphics[width=0.95\textwidth]{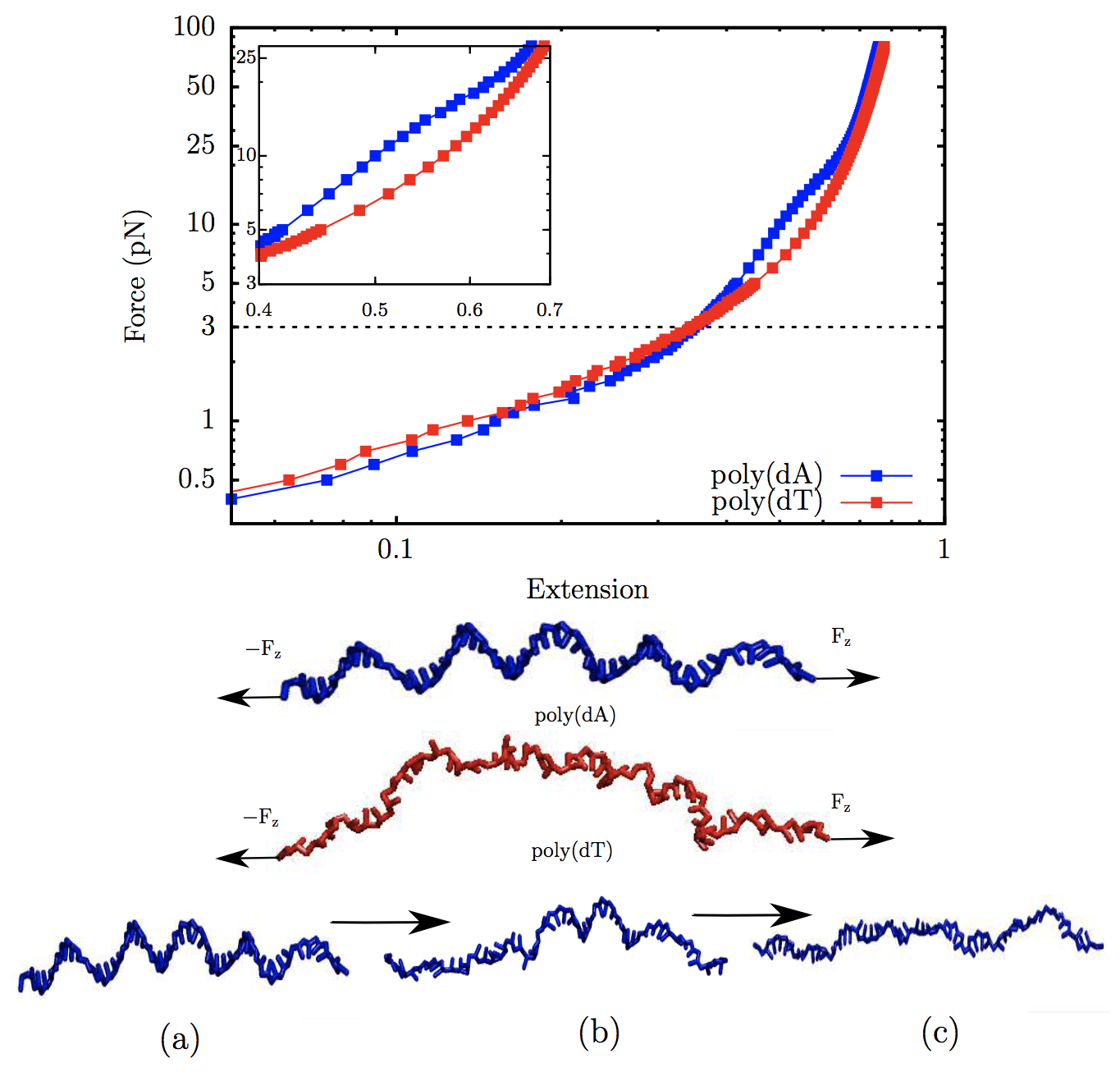}
\caption{Top: The force extension behavior of ssDNA.  The profiles correspond to a salt concentration of 0.5\,M. The z-axis denotes the extension, $z$, normalized by the contour length, $L$. The red curve denotes the force extension profile for poly(dT). The blue curve denotes the profile for poly(dA). A distinct plateau appears in the force-extension profile due to helix-coil transition (see inset). In contrast, the force-extension curve for poly(dT) follows the conventional worm-like chain behavior. The dashed line corresponds to the critical force ($\sim$ 3 pN) at which the extension of the poly(dT) chain exceeds that of poly(dA). 
Middle: Representative snaphots of the poly(dA) and poly(dT) sequences at 3\,pN. The arrows represent the direction of the applied force. While poly(dA) consists of stacked helical domains, poly(dT) is mostly unstacked. 
 Bottom: Helix-to-coil transition in the poly(dA) sequence that results in a plateau in the force-extension profile. In snapshot (a) the strand is under a tension of 3\,pN, and helical domains persist throughout the chain. Snapshot (b) corresponds to a tension of 17\,pN, where approximately two helical domains remain. In snapshot (c), the strand experiences a tension of 30\,pN, and no visible helical domains exist in the chain.}
\end{figure}

To investigate how the TIS-DNA model captures sequence-specific effects on the force-extension behavior of ssDNA, we simulated the mechanical stretching of poly(dA) and poly(dT) strands consisting of 50 nucleotides, at a salt concentration of 0.5\,M.  The force-extension curves are depicted in Figure 11.  While the mechanical response of poly(dT) is purely entropic,  the force-extension profile of poly(dA) exhibits a concave feature between $\sim$7 and 22\,pN, which corresponds to the plateau reported in experiments.\cite{fex1,fex2,fex3}  A substantial fraction of bases in poly(dA) are stacked, and form helical domains, at forces below $\sim$7\,pN.  At higher forces, a helix-to-coil transition (Figure 11) unravels the helical domains. At low forces, the largely unstacked poly(dT) sequence typically has a shorter extension, compared to poly(dA),  as it is more flexible and has a propensity to collapse. On the other hand, the poly(dA) strand is more extensible in this force regime because stacked helical domains are associated with a smaller entropic cost of aligning in the direction of the applied force.  The strands align with the force with greater ease, as the force increases, and the curves cross at around $\sim$ 3\,pN.  The critical force for crossover is in excellent agreement with the experimental estimate ($\sim$ 4\,pN) of Saleh and coworkers.\cite{fex3}

\subsection{Stacking thermodynamics of ssDNA}

To assess whether the TIS-DNA model provides a robust description of stacking thermodynamics, over a wide range of temperature, we consider a 14 nucleotide long ssDNA with the sequence, 5$^{\prime}$GCGTCATACAGTGC3$^{\prime}$, for which experimental data is available from Holbrook \textit{et al.}\cite{holbrok} In the experiment, stacking probability is described in terms of relative absorbance,  with unstacked regions showing a higher absorbance compared to stacked bases.  In our model, we consider a dimer to be stacked if $U_{S} < -k_{B}T$. The average stacking probability, $\langle p_{stack} \rangle$,  as a function of temperature, for the ssDNA sequence is shown in Figure 12.  As expected, $\langle p_{stack} \rangle$ decreases linearly with temperature.  Our estimates lie within the error bars of the experimental values, particularly in the low temperature regime.  

\begin{figure}
\includegraphics[width=0.95\textwidth]{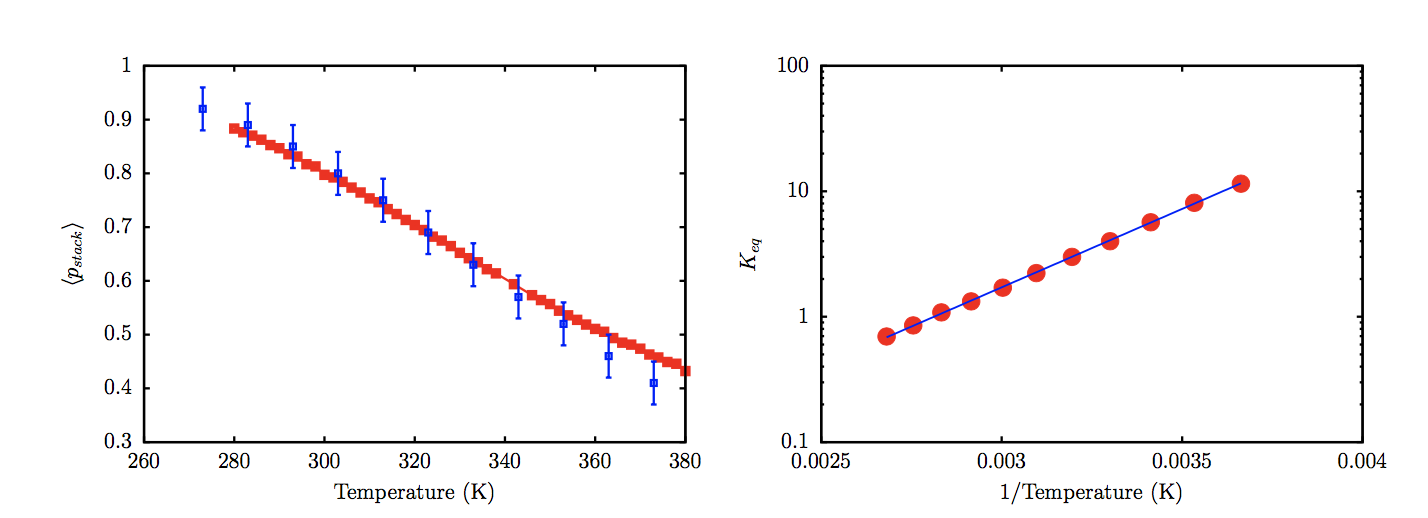}
\caption{Left: The red curve shows the evolution of stacking probability, $\langle p_{stack} \rangle$, with temperature. The blue squares are experimental data from Holbrook \textit{et al.}\cite{holbrok}.  The simulation, and experimental data correspond to a salt concentration of 0.12\,M. Right: A van't Hoff plot depicting the variation of $K_{eq}$ with $T$. The red circles are the estimated equilibrium constants from simulations, and the blue solid line denotes a linear fit. From the fit, we estimate $\Delta H_{stack} = -4.8$\,kcal/mol, $\Delta S_{stack} = -13.3$\,cal/mol/K, and $T_{m} = 363$\,K. These values are in very good agreement with the corresponding experimental estimates, which are: $\Delta H_{stack} = -5.7$\,kcal/mol; $\Delta S = -16.0$\,cal/mol/K; $T_{m} = 356$\,K.}
\end{figure}

Assuming a two-state model, we can define the equilibrium constant, $K_{eq}$ for stacking in terms of stacking probability:
\begin{equation}
K_{eq} = \frac{\langle p_{stack} \rangle}{1 - \langle p_{stack} \rangle }
\end{equation}

\noindent

A van't Hoff analysis (Figure 11)  based on equation (21), yields stacking enthalpy, $\Delta H_{stack} = -4.8$\,kcal/mol, entropy, $\Delta S_{stack} = -13.3$\,cal/mol/K, and the transition midpoint temperature $T_{m} = 363$\,K.  These values are in good agreement with those reported by Holbrook \textit{et al.} at the same salt concentration (see Figure 12).



\subsection{Elasticity of double-stranded DNA}





To assess the accuracy of the TIS-DNA model in describing the elasticity of dsDNA, we consider two DNA sequences of length 60 and 130 base pairs, considered in an earlier study.\cite{CG_plotkin} The sequences of the leading strands for the two sequences are: 

\noindent
\textit{Seq1:}  5$^{\prime}$-CATCCTCGACAATCGGAACCAGGAAGCGCCCCGCAACTCTGCCGCGATCG
\newline 
GTGTTCGCCT-3$^{\prime}$
\newline
\textit{Seq2:} 5$^{\prime}$-GCATCCTCGACAATCGGAACCAGGAAGCGCCCCGCAACTCTGCCGCGATCG
\newline
GTGTTCGCCTCCAAGCTAGAACCTGGCGATACGGCCTAAGGGCTCCGGAACAAGC
\newline
TGAGGCCTTGGCCGTTTAAGGCCG-3$^{\prime}$

Compared to ssDNA, the decay of the tangent autocorrelations are exponential, and dsDNA exhibits a worm like chain behavior over the entire range of salt concentration \mbox{(Figure 13)}.  Hence, equation (18) was used to compute the persistence lengths. \cite{Baumann1997,smith_science}  
\begin{figure}
\includegraphics[width=0.95\textwidth]{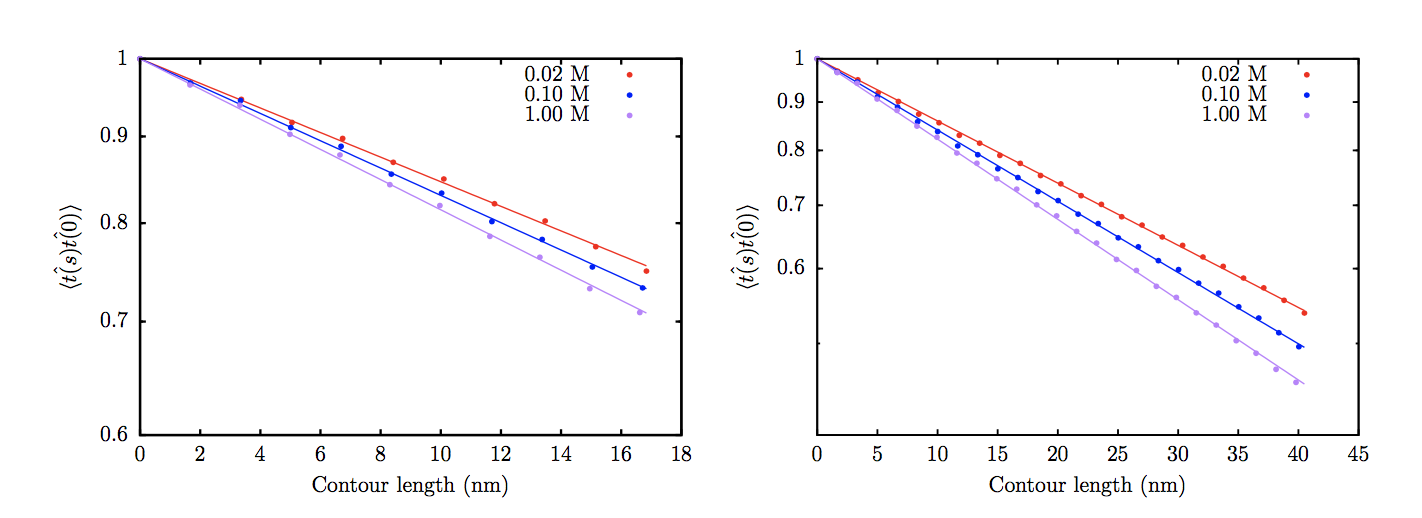}
\caption{The decay of the tangent correlations with the contour length in \textit{Seq1} (left), and \textit{Seq2} (right), at three different salt concentrations.  The correlations decay  exponentially in contrast to ssDNA.}
\end{figure}

\begin{figure}
\includegraphics[width=0.95\textwidth]{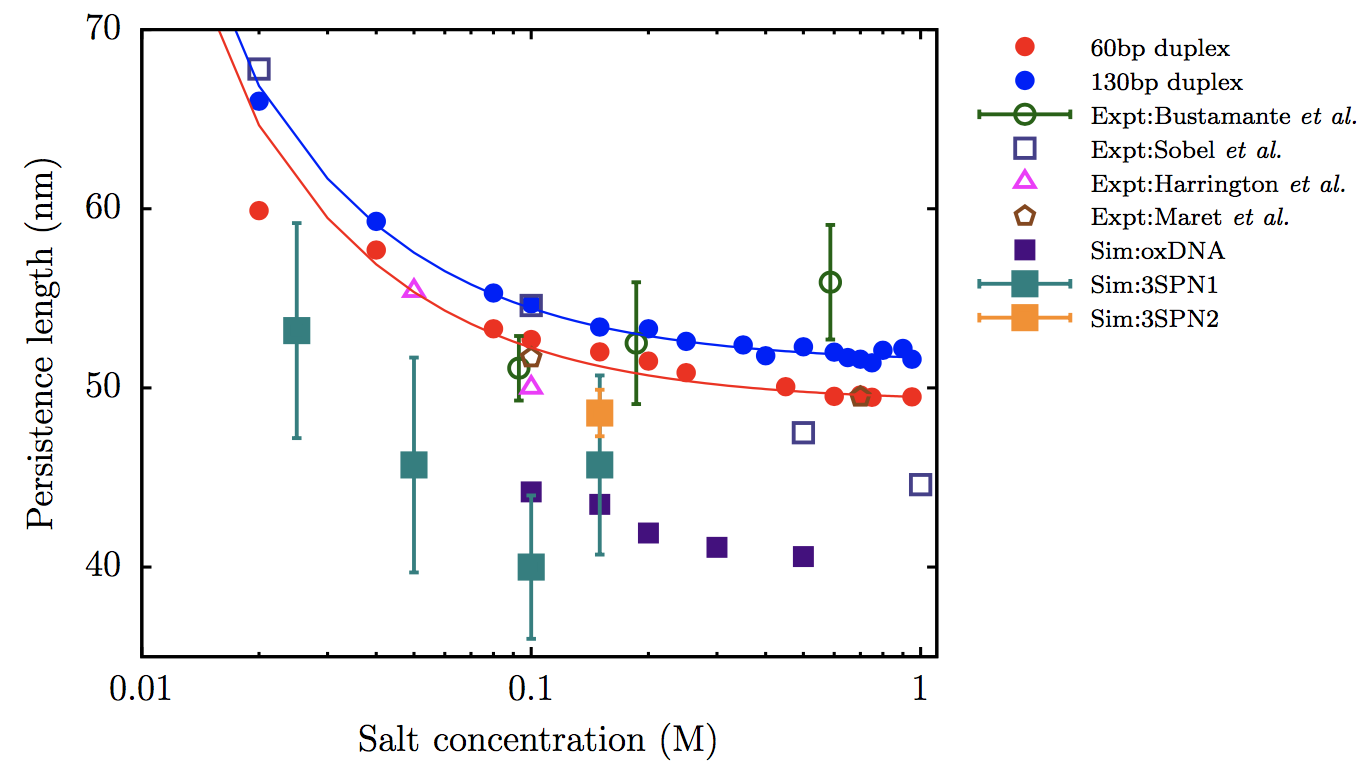}
\caption{The variation of persistence length with salt concentration for the 60 bp duplex, \textit{Seq1} (red) and 130 bp duplex, \textit{Seq2} (blue) from simulations.  The solid curves denote fits of the simulation data to the OSF theory. The green open circles are $l_{p}$ values reported by Bustamante and coworkers.\cite{Baumann1997}.  The triangles denote experimental data of Harrington \textit{et al.}\cite{harrington_lp} The open squares are data from the experiments of  Sobel and Harpst\cite{sobel_lp} on bacteriophage DNA. The brown polygon corresponds to the measurement of Maret and Weill.\cite{maret_lp} The filled squares denote the persistence lengths estimated by other groups: 3SPN1 (teal),\cite{Sambriski2009} 3SPN2 (orange),\cite{depablo_3spn2} oxDNA (purple).\cite{oxDNA2} We do not show the data from Plotkin and coworkers,\cite{CG_plotkin} because the estimated $l_{p}$ values are too low, and lie outside the experimentally reported range.}
\end{figure}

As expected, the presence of hydrogen-bonding interactions ($U_{HB}$) between complementary strands induces additional bending rigidity. 
For both the dsDNA sequences, the $l_{p}$ values predicted by the TIS-DNA model are in good agreement with the various experimental estimates.\cite{Baumann1997,harrington_lp,sobel_lp,maret_lp}. In particular, we obtain very good agreement with the data of Bustamante and coworkers\cite{Baumann1997} over the ion concentration range spanning an order of magnitude.  At low salt concentrations, our $l_{p}$ estimate for \textit{Seq2} is close to the value reported by Sobel and Harpst for bacteriophage DNA.\cite{sobel_lp}. In the regime corresponding to moderate salt concentration, our results are in very good agreement with the values reported in separate studies by Harrington \textit{et al.},\cite{harrington_lp} and Maret and Weill.\cite{maret_lp}

For both \textit{Seq1}, and \textit{Seq2}, the variation of $l_{p}$ with salt concentration is in accord with the OSF theory.  A fit to equation (24) yields a bare persistence length, $l_{p0}$, of 49.3\,nm for \textit{Seq1} and 51.5\,nm for \textit{Seq2}. These estimates fall within the range suggested by Bustamante and coworkers.\cite{Baumann1997}
Although sequence-dependent variations in $l_{p}$ are to be expected in general,\cite{maddocks,volgod} the results predicted by the TIS-DNA model already look promising, considering that we did not parametrize the model explicitly to reproduce the dsDNA persistence lengths.  





\section{Conclusions}

In this work, we have introduced a robust coarse-grained model of DNA based on the TIS representation of nucleic acids, which reproduces the sequence-dependent mechanical properties of both single-stranded, and double-stranded DNA. The model represents a significant improvement over current coarse-grained DNA models, particularly in the description of single-stranded DNA flexibility.  In particular, we are able to reproduce experimental trends in sequence and salt-dependent persistence lengths, Flory scaling exponents, and force-extension behavior.  Once the various interaction strengths are optimized for ssDNA, an appropriate choice of a single parameter ($U_{HB}$) is able to reproduce the melting profile of a DNA hairpin, as well as the persistence length of dsDNA.  Due to its balanced description of both single and double-stranded DNA,  the TIS-DNA model in its current form should be well suited for providing the much needed molecular insight into folding thermodynamics, duplex association, as well as force extension behavior of dsDNA.  Our parametrization strategy is quite general, and we envisage many other novel applications of the TIS-DNA model in problems of contemporary interest, including DNA self-assembly and material design, as well as the study of DNA-protein, and DNA-RNA interactions. 

The current version of the model does not include counterions explicitly, and only provides a description of electrostatics at the Debye-H{\"u}ckel level.  The possibility of non-native hydrogen bonding, and stacking, which could be a key determinant in many important processes involving DNA is excluded from the model. Recently, such extensions have been included in the RNA version of the Three Interaction Site model, and it dramatically improves the description of RNA folding thermodynamics, and assembly.\cite{Denesyuk_nature}  These avenues will be further explored in future work.

\begin{acknowledgement}
We are grateful to Upayan Baul, Hung Nguyen and Huong Vu for fruitful discussions. This research was supported by the National Science Foundation (CHE 16-36424), and the Collie-Welch Regents Chair (F0019).
\end{acknowledgement}












\newpage


\begin{mcitethebibliography}{93}
\providecommand*\natexlab[1]{#1}
\providecommand*\mciteSetBstSublistMode[1]{}
\providecommand*\mciteSetBstMaxWidthForm[2]{}
\providecommand*\mciteBstWouldAddEndPuncttrue
  {\def\EndOfBibitem{\unskip.}}
\providecommand*\mciteBstWouldAddEndPunctfalse
  {\let\EndOfBibitem\relax}
\providecommand*\mciteSetBstMidEndSepPunct[3]{}
\providecommand*\mciteSetBstSublistLabelBeginEnd[3]{}
\providecommand*\EndOfBibitem{}
\mciteSetBstSublistMode{f}
\mciteSetBstMaxWidthForm{subitem}{(\alph{mcitesubitemcount})}
\mciteSetBstSublistLabelBeginEnd
  {\mcitemaxwidthsubitemform\space}
  {\relax}
  {\relax}

\bibitem[Saenger(1984)]{sanger}
Saenger,~W. \emph{Principles of Nucleic Acid Structure}; Springer-Verlag,
  Berlin, 1984\relax
\mciteBstWouldAddEndPuncttrue
\mciteSetBstMidEndSepPunct{\mcitedefaultmidpunct}
{\mcitedefaultendpunct}{\mcitedefaultseppunct}\relax
\EndOfBibitem
\bibitem[Reif \latin{et~al.}(1999)Reif, Clausen-Schaumann, and
  Gaub]{seq_mechanics1}
Reif,~M.; Clausen-Schaumann,~H.; Gaub,~H.~E. Sequence-dependent mechanics of
  single DNA molecules. \emph{Nat. Struct. Biol.} \textbf{1999}, \emph{6},
  346--349\relax
\mciteBstWouldAddEndPuncttrue
\mciteSetBstMidEndSepPunct{\mcitedefaultmidpunct}
{\mcitedefaultendpunct}{\mcitedefaultseppunct}\relax
\EndOfBibitem
\bibitem[Bonev and Cavalli(2016)Bonev, and Cavalli]{chromatin_organisation}
Bonev,~B.; Cavalli,~G. Organization and function of the 3D genome. \emph{Nat.
  Rev. Genetics} \textbf{2016}, \emph{17}, 661--678\relax
\mciteBstWouldAddEndPuncttrue
\mciteSetBstMidEndSepPunct{\mcitedefaultmidpunct}
{\mcitedefaultendpunct}{\mcitedefaultseppunct}\relax
\EndOfBibitem
\bibitem[Seeman(2010)]{dna_nanotech2}
Seeman,~N.~C. Nanomaterials based on DNA. \emph{Annu. Rev. Biochem.}
  \textbf{2010}, \emph{79}, 65--87\relax
\mciteBstWouldAddEndPuncttrue
\mciteSetBstMidEndSepPunct{\mcitedefaultmidpunct}
{\mcitedefaultendpunct}{\mcitedefaultseppunct}\relax
\EndOfBibitem
\bibitem[Chen \latin{et~al.}(2015)Chen, Groves, Muscat, and
  Selig]{dna_nanotech1}
Chen,~Y.~J.; Groves,~B.; Muscat,~R.~A.; Selig,~G. DNA nanotechnology from the
  test tube to the cell. \emph{Nature Nanotechnol.} \textbf{2015}, \emph{10},
  748--760\relax
\mciteBstWouldAddEndPuncttrue
\mciteSetBstMidEndSepPunct{\mcitedefaultmidpunct}
{\mcitedefaultendpunct}{\mcitedefaultseppunct}\relax
\EndOfBibitem
\bibitem[Beveridge \latin{et~al.}(2004)Beveridge, Barreiro, Byun, Case,
  Cheatham, Dixit, Giudice, Lankas, Lavery, Maddocks, Osman, Seibert, Sklenar,
  Stoll, Thayer, Varnai, and Young]{Beveridge2004}
Beveridge,~D.~L.; Barreiro,~G.; Byun,~K.~S.; Case,~D.~A.; Cheatham,~T.~E.;
  Dixit,~S.~B.; Giudice,~E.; Lankas,~F.; Lavery,~R.; Maddocks,~J.~H.;
  Osman,~R.; Seibert,~E.; Sklenar,~H.; Stoll,~G.; Thayer,~K.~M.; Varnai,~P.;
  Young,~M.~A. {Molecular Dynamics Simulations of the 136 Unique
  Tetranucleotide Sequences of DNA Oligonucleotides. I. Research Design and
  Results on d(C(p)G) Steps}. \emph{Biophys. J.} \textbf{2004}, \emph{87},
  3799--3813\relax
\mciteBstWouldAddEndPuncttrue
\mciteSetBstMidEndSepPunct{\mcitedefaultmidpunct}
{\mcitedefaultendpunct}{\mcitedefaultseppunct}\relax
\EndOfBibitem
\bibitem[Lavery \latin{et~al.}(2010)Lavery, Zakrzewska, Beveridge, Bishop,
  Case, Cheatham, Dixit, Jayaram, Lankas, Laughton, Maddocks, Michon, Osman,
  Orozco, Perez, Singh, Spackova, and Sponer]{Lavery2010}
Lavery,~R.; Zakrzewska,~K.; Beveridge,~D.; Bishop,~T.~C.; Case,~D.~A.;
  Cheatham,~T.; Dixit,~S.; Jayaram,~B.; Lankas,~F.; Laughton,~C.;
  Maddocks,~J.~H.; Michon,~A.; Osman,~R.; Orozco,~M.; Perez,~A.; Singh,~T.;
  Spackova,~N.; Sponer,~J. {A systematic molecular dynamics study of
  nearest-neighbor effects on base pair and base pair step conformations and
  fluctuations in B-DNA}. \emph{Nucleic Acids Res.} \textbf{2010}, \emph{38},
  299--313\relax
\mciteBstWouldAddEndPuncttrue
\mciteSetBstMidEndSepPunct{\mcitedefaultmidpunct}
{\mcitedefaultendpunct}{\mcitedefaultseppunct}\relax
\EndOfBibitem
\bibitem[Hyeon and Thirumalai(2011)Hyeon, and Thirumalai]{Hyeon11NatComm}
Hyeon,~C.; Thirumalai,~D. Capturing the essence of folding and functions of
  biomolecules using coarse-grained models. \emph{Nat. Comm.} \textbf{2011},
  \emph{2}, 1481\relax
\mciteBstWouldAddEndPuncttrue
\mciteSetBstMidEndSepPunct{\mcitedefaultmidpunct}
{\mcitedefaultendpunct}{\mcitedefaultseppunct}\relax
\EndOfBibitem
\bibitem[Chen \latin{et~al.}(2010)Chen, Darst, and Thirumalai]{Chen09PNAS}
Chen,~J.; Darst,~S.~A.; Thirumalai,~D. Promoter melting triggered by bacterial
  RNA polymerase occurs in three steps. \emph{Proc. Natl. Acad. Sci. USA}
  \textbf{2010}, \emph{107}, 12523--12528\relax
\mciteBstWouldAddEndPuncttrue
\mciteSetBstMidEndSepPunct{\mcitedefaultmidpunct}
{\mcitedefaultendpunct}{\mcitedefaultseppunct}\relax
\EndOfBibitem
\bibitem[Fosado \latin{et~al.}(2016)Fosado, Michieletto, Allan, Brackley,
  Henrich, and Marenduzzo]{Merenduzzo17SoftMatter}
Fosado,~Y. A.~G.; Michieletto,~D.; Allan,~J.; Brackley,~C.; Henrich,~O.;
  Marenduzzo,~D. A single nucleotide resolution model for large-scale
  simulations of double stranded DNA. \emph{Soft Matter} \textbf{2016},
  \emph{47}, 9458--9470\relax
\mciteBstWouldAddEndPuncttrue
\mciteSetBstMidEndSepPunct{\mcitedefaultmidpunct}
{\mcitedefaultendpunct}{\mcitedefaultseppunct}\relax
\EndOfBibitem
\bibitem[{Hyeon, C. and Thirumalai}(2005)]{HyeonC.andThirumalai2005}
{Hyeon, C. and Thirumalai},~D. {Mechanical unfolding of RNA hairpins}.
  \emph{Proc. Nat. Acad. Sci. U.S.A.} \textbf{2005}, \emph{102},
  6789--6794\relax
\mciteBstWouldAddEndPuncttrue
\mciteSetBstMidEndSepPunct{\mcitedefaultmidpunct}
{\mcitedefaultendpunct}{\mcitedefaultseppunct}\relax
\EndOfBibitem
\bibitem[Moriss-Andrews \latin{et~al.}(2010)Moriss-Andrews, Rotler, and
  Plotkin]{CG_plotkin}
Moriss-Andrews,~A.; Rotler,~J.; Plotkin,~S.~S. A systematically coarse-grained
  model for DNA and its predictions for persistence length, stacking, twist,
  and chirality. \emph{J. Chem. Phys.} \textbf{2010}, \emph{132}, 035105\relax
\mciteBstWouldAddEndPuncttrue
\mciteSetBstMidEndSepPunct{\mcitedefaultmidpunct}
{\mcitedefaultendpunct}{\mcitedefaultseppunct}\relax
\EndOfBibitem
\bibitem[Ouldridge \latin{et~al.}(2011)Ouldridge, Louis, and Doye]{oxDNA}
Ouldridge,~T.~E.; Louis,~A.~A.; Doye,~J. P.~K. Structural, mechanical, and
  thermodynamic properties of a coarse-grained DNA model. \emph{J. Chem. Phys.}
  \textbf{2011}, \emph{134}, 085101\relax
\mciteBstWouldAddEndPuncttrue
\mciteSetBstMidEndSepPunct{\mcitedefaultmidpunct}
{\mcitedefaultendpunct}{\mcitedefaultseppunct}\relax
\EndOfBibitem
\bibitem[Hinckley \latin{et~al.}(2013)Hinckley, Freeman, Whitmer, and
  de~Pablo]{depablo_3spn2}
Hinckley,~D.~M.; Freeman,~G.~S.; Whitmer,~J.~K.; de~Pablo,~J.~J. An
  experimentally-informed coarse-grained 3-Site-Per-Nucleotide model of DNA:
  structure, thermodynamics, and dynamics of hybridization. \emph{J. Chem.
  Phys.} \textbf{2013}, \emph{139}, 144903\relax
\mciteBstWouldAddEndPuncttrue
\mciteSetBstMidEndSepPunct{\mcitedefaultmidpunct}
{\mcitedefaultendpunct}{\mcitedefaultseppunct}\relax
\EndOfBibitem
\bibitem[Maciejczyk \latin{et~al.}(2014)Maciejczyk, Spasic, Liwo, and
  Scheraga]{Maciejczyk2014}
Maciejczyk,~M.; Spasic,~A.; Liwo,~A.; Scheraga,~H.~A. {DNA duplex formation
  with a coarse-grained model}. \emph{J. Chem. Theory Comput.} \textbf{2014},
  \emph{10}, 5020--5035\relax
\mciteBstWouldAddEndPuncttrue
\mciteSetBstMidEndSepPunct{\mcitedefaultmidpunct}
{\mcitedefaultendpunct}{\mcitedefaultseppunct}\relax
\EndOfBibitem
\bibitem[Uusitalo \latin{et~al.}(2015)Uusitalo, Ing{\'{o}}lfsson, Akhshi,
  Tieleman, and Marrink]{Uusitalo2015}
Uusitalo,~J.~J.; Ing{\'{o}}lfsson,~H.~I.; Akhshi,~P.; Tieleman,~D.~P.;
  Marrink,~S.~J. {Martini Coarse-Grained Force Field: Extension to DNA}.
  \emph{J. Chem. Theory Comput.} \textbf{2015}, \emph{11}, 3932--3945\relax
\mciteBstWouldAddEndPuncttrue
\mciteSetBstMidEndSepPunct{\mcitedefaultmidpunct}
{\mcitedefaultendpunct}{\mcitedefaultseppunct}\relax
\EndOfBibitem
\bibitem[Cho \latin{et~al.}(2009)Cho, Pincus, and Thirumalai]{Cho09PNAS}
Cho,~S.~S.; Pincus,~D.~L.; Thirumalai,~D. Assembly mechanisms of RNA
  pseudoknots are determined by the stabilities of constituent secondary
  structures. \emph{Proc. Natl. Acad. Sci. USA} \textbf{2009}, \emph{106},
  17349--17354\relax
\mciteBstWouldAddEndPuncttrue
\mciteSetBstMidEndSepPunct{\mcitedefaultmidpunct}
{\mcitedefaultendpunct}{\mcitedefaultseppunct}\relax
\EndOfBibitem
\bibitem[Savelyev and Papoian(2010)Savelyev, and Papoian]{papoian_CG}
Savelyev,~A.; Papoian,~G. Chemically accurate coarse graining of
  double-stranded DNA. \emph{Proc. Natl. Acad. Sci. USA} \textbf{2010},
  \emph{107}, 20340--20345\relax
\mciteBstWouldAddEndPuncttrue
\mciteSetBstMidEndSepPunct{\mcitedefaultmidpunct}
{\mcitedefaultendpunct}{\mcitedefaultseppunct}\relax
\EndOfBibitem
\bibitem[Markegard \latin{et~al.}(2015)Markegard, Fu, Reddy, and
  Nguyen]{CG_ngyen}
Markegard,~C.~B.; Fu,~I.~W.; Reddy,~A.; Nguyen,~H.~D. Coarse-Grained Simulation
  Study of Sequence Effects on DNA Hybridization in a Concentrated Environment.
  \emph{J. Phys. Chem. B} \textbf{2015}, \emph{119}, 1823--1834\relax
\mciteBstWouldAddEndPuncttrue
\mciteSetBstMidEndSepPunct{\mcitedefaultmidpunct}
{\mcitedefaultendpunct}{\mcitedefaultseppunct}\relax
\EndOfBibitem
\bibitem[Brini \latin{et~al.}(2013)Brini, Algaer, Ganguly, Li,
  Rodriguez-Ropero, and {van der Vegt}]{CG_review}
Brini,~E.; Algaer,~E.~A.; Ganguly,~P.; Li,~C.; Rodriguez-Ropero,~F.; {van der
  Vegt},~N. F.~A. Systematic coarse-graining methods for soft matter
  simulations – a review. \emph{Soft Matter} \textbf{2013}, \emph{9},
  2108--2119\relax
\mciteBstWouldAddEndPuncttrue
\mciteSetBstMidEndSepPunct{\mcitedefaultmidpunct}
{\mcitedefaultendpunct}{\mcitedefaultseppunct}\relax
\EndOfBibitem
\bibitem[Drukker \latin{et~al.}(2001)Drukker, Wu, and Schatz]{Drukker2000}
Drukker,~K.; Wu,~G.; Schatz,~G.~C. {Model simulations of DNA denaturation
  dynamics}. \emph{J. Chem. Phys.} \textbf{2001}, \emph{114}, 579--590\relax
\mciteBstWouldAddEndPuncttrue
\mciteSetBstMidEndSepPunct{\mcitedefaultmidpunct}
{\mcitedefaultendpunct}{\mcitedefaultseppunct}\relax
\EndOfBibitem
\bibitem[Dans \latin{et~al.}(2010)Dans, Zeida, MacHado, and Pantano]{Dans2010}
Dans,~P.~D.; Zeida,~A.; MacHado,~M.~R.; Pantano,~S. {A coarse grained model for
  atomic-detailed DNA simulations with explicit electrostatics}. \emph{J. Chem.
  Theory Comput.} \textbf{2010}, \emph{6}, 1711--1725\relax
\mciteBstWouldAddEndPuncttrue
\mciteSetBstMidEndSepPunct{\mcitedefaultmidpunct}
{\mcitedefaultendpunct}{\mcitedefaultseppunct}\relax
\EndOfBibitem
\bibitem[Sambriski \latin{et~al.}(2009)Sambriski, Schwartz, and
  de~Pablo]{Sambriski2009}
Sambriski,~E.; Schwartz,~D.; de~Pablo,~J.~J. {A Mesoscale Model of DNA and Its
  Renaturation}. \emph{Biophys. J.} \textbf{2009}, \emph{96}, 1675--1690\relax
\mciteBstWouldAddEndPuncttrue
\mciteSetBstMidEndSepPunct{\mcitedefaultmidpunct}
{\mcitedefaultendpunct}{\mcitedefaultseppunct}\relax
\EndOfBibitem
\bibitem[Denesyuk and Thirumalai(2013)Denesyuk, and Thirumalai]{Denesyuk2013}
Denesyuk,~N.~A.; Thirumalai,~D. {A Coarse-Grained Model for Predicting RNA
  Folding Thermodynamics.} \emph{J. Phys. Chem. B} \textbf{2013}, \emph{117},
  4901--4911\relax
\mciteBstWouldAddEndPuncttrue
\mciteSetBstMidEndSepPunct{\mcitedefaultmidpunct}
{\mcitedefaultendpunct}{\mcitedefaultseppunct}\relax
\EndOfBibitem
\bibitem[Denesyuk and Thirumalai(2011)Denesyuk, and Thirumalai]{crowding_TIS}
Denesyuk,~N.~A.; Thirumalai,~D. Crowding Promotes the Switch from Hairpin to
  Pseudoknot Conformation in Human Telomerase RNA. \emph{J. Am. Chem. Soc.}
  \textbf{2011}, \emph{133}, 11858--11861\relax
\mciteBstWouldAddEndPuncttrue
\mciteSetBstMidEndSepPunct{\mcitedefaultmidpunct}
{\mcitedefaultendpunct}{\mcitedefaultseppunct}\relax
\EndOfBibitem
\bibitem[Moore \latin{et~al.}(2014)Moore, Iacovella, and McCabe]{IBI}
Moore,~T.~C.; Iacovella,~C.~R.; McCabe,~C. Derivation of coarse-grained
  potentials via multistate iterative Boltzmann inversion. \emph{J. Chem.
  Phys.} \textbf{2014}, \emph{140}, 224104\relax
\mciteBstWouldAddEndPuncttrue
\mciteSetBstMidEndSepPunct{\mcitedefaultmidpunct}
{\mcitedefaultendpunct}{\mcitedefaultseppunct}\relax
\EndOfBibitem
\bibitem[Xia \latin{et~al.}(2010)Xia, Gardner, Gutell, and Ren]{Xia2010}
Xia,~Z.; Gardner,~D.~P.; Gutell,~R.~R.; Ren,~P. {Coarse-grained model for
  simulation of RNA three-dimensional structures}. \emph{J Phys Chem B}
  \textbf{2010}, \emph{114}, 13497--13506\relax
\mciteBstWouldAddEndPuncttrue
\mciteSetBstMidEndSepPunct{\mcitedefaultmidpunct}
{\mcitedefaultendpunct}{\mcitedefaultseppunct}\relax
\EndOfBibitem
\bibitem[Agrawal \latin{et~al.}(2014)Agrawal, Arya, and Oswald]{Agrawal2014}
Agrawal,~V.; Arya,~G.; Oswald,~J. {Simultaneous iterative boltzmann inversion
  for coarse-graining of polyurea}. \emph{Macromolecules} \textbf{2014},
  \emph{47}, 3378--3389\relax
\mciteBstWouldAddEndPuncttrue
\mciteSetBstMidEndSepPunct{\mcitedefaultmidpunct}
{\mcitedefaultendpunct}{\mcitedefaultseppunct}\relax
\EndOfBibitem
\bibitem[Chandler \latin{et~al.}(1983)Chandler, Weeks, and Andersen]{WCA}
Chandler,~D.; Weeks,~J.~D.; Andersen,~H.~C. Van der waals picture of liquids,
  solids, and phase transformations. \emph{Science} \textbf{1983}, \emph{220},
  787--794\relax
\mciteBstWouldAddEndPuncttrue
\mciteSetBstMidEndSepPunct{\mcitedefaultmidpunct}
{\mcitedefaultendpunct}{\mcitedefaultseppunct}\relax
\EndOfBibitem
\bibitem[Dima \latin{et~al.}(2005)Dima, Hyeon, and Thirumalai]{stacking_Dima}
Dima,~R.~I.; Hyeon,~C.; Thirumalai,~D. Extracting stacking interaction
  parameters for RNA from the data set of native structures. \emph{J. Mol.
  Biol.} \textbf{2005}, \emph{347}, 53--69\relax
\mciteBstWouldAddEndPuncttrue
\mciteSetBstMidEndSepPunct{\mcitedefaultmidpunct}
{\mcitedefaultendpunct}{\mcitedefaultseppunct}\relax
\EndOfBibitem
\bibitem[Santalucia \latin{et~al.}(1996)Santalucia, Allawi, and
  Seneviratne]{nearest}
Santalucia,~J.; Allawi,~H.~T.; Seneviratne,~P.~A. Improved nearest-neighbor
  parameters for predicting DNA duplex stability. \emph{Biochemistry}
  \textbf{1996}, \emph{35}, 3555--3562\relax
\mciteBstWouldAddEndPuncttrue
\mciteSetBstMidEndSepPunct{\mcitedefaultmidpunct}
{\mcitedefaultendpunct}{\mcitedefaultseppunct}\relax
\EndOfBibitem
\bibitem[Santalucia and Hicks(2004)Santalucia, and Hicks]{Santalucia2004}
Santalucia,~J.; Hicks,~D. {The thermodynamics of DNA structural motifs}.
  \emph{Annu Rev Biophys Biomol Struct} \textbf{2004}, \emph{33},
  415--440\relax
\mciteBstWouldAddEndPuncttrue
\mciteSetBstMidEndSepPunct{\mcitedefaultmidpunct}
{\mcitedefaultendpunct}{\mcitedefaultseppunct}\relax
\EndOfBibitem
\bibitem[Yakovchuk \latin{et~al.}(2006)Yakovchuk, Protozanova, and
  Frank-Kamnetskii]{kamenetskii}
Yakovchuk,~P.; Protozanova,~E.; Frank-Kamnetskii,~M.~D. Base-stacking and
  base-pairing contributions into thermal stability of the DNA double helix.
  \emph{Nucleic Acids Res.} \textbf{2006}, \emph{34}, 564--574\relax
\mciteBstWouldAddEndPuncttrue
\mciteSetBstMidEndSepPunct{\mcitedefaultmidpunct}
{\mcitedefaultendpunct}{\mcitedefaultseppunct}\relax
\EndOfBibitem
\bibitem[Olsthoorn \latin{et~al.}(1981)Olsthoorn, Bostelaar, {De Rooij}, {Van
  Boom}, and Altona]{OLSTHOORN1981}
Olsthoorn,~S.; Bostelaar,~L.~J.; {De Rooij},~J.~F.; {Van Boom},~J.~H.;
  Altona,~C. {Circular Dichroism Study of Stacking Properties of
  Oligodeoxyadenylates and Polydeoxyadenylate: A Three‐State Conformational
  Model}. \emph{Eur. J. Biochem.} \textbf{1981}, \emph{115}, 309--321\relax
\mciteBstWouldAddEndPuncttrue
\mciteSetBstMidEndSepPunct{\mcitedefaultmidpunct}
{\mcitedefaultendpunct}{\mcitedefaultseppunct}\relax
\EndOfBibitem
\bibitem[Solie and Schellman(1968)Solie, and Schellman]{sollie}
Solie,~T.~N.; Schellman,~J.~A. The interaction of nucleosides in aqueous
  solution. \emph{J. Mol. Biol} \textbf{1968}, \emph{33}, 61--77\relax
\mciteBstWouldAddEndPuncttrue
\mciteSetBstMidEndSepPunct{\mcitedefaultmidpunct}
{\mcitedefaultendpunct}{\mcitedefaultseppunct}\relax
\EndOfBibitem
\bibitem[Flori{\'{a}}n \latin{et~al.}(1999)Flori{\'{a}}n, {\v S}poner, and
  Warshel]{sponer}
Flori{\'{a}}n,~J.; {\v S}poner,~J.; Warshel,~A. Thermodynamic Parameters for
  Stacking and Hydrogen Bonding of Nucleic Acid Bases in Aqueous Solution: 
  Ab Initio/Langevin Dipoles Study. \emph{J. Phys. Chem. B} \textbf{1999},
  \emph{103}, 884--892\relax
\mciteBstWouldAddEndPuncttrue
\mciteSetBstMidEndSepPunct{\mcitedefaultmidpunct}
{\mcitedefaultendpunct}{\mcitedefaultseppunct}\relax
\EndOfBibitem
\bibitem[Jafilan \latin{et~al.}(2012)Jafilan, Klein, Hyun, and
  Flori{\'{a}}n]{Jafilan2012}
Jafilan,~S.; Klein,~L.; Hyun,~C.; Flori{\'{a}}n,~J. {Intramolecular base
  stacking of dinucleoside monophosphate anions in aqueous solution}. \emph{J.
  Phys. Chem. B} \textbf{2012}, \emph{116}, 3613--3618\relax
\mciteBstWouldAddEndPuncttrue
\mciteSetBstMidEndSepPunct{\mcitedefaultmidpunct}
{\mcitedefaultendpunct}{\mcitedefaultseppunct}\relax
\EndOfBibitem
\bibitem[Brown \latin{et~al.}(2015)Brown, Andrews, and Elcock]{brown_jctc}
Brown,~R.~F.; Andrews,~C.~T.; Elcock,~A.~H. Stacking Free Energies of All DNA
  and RNA Nucleoside Pairs and Dinucleoside-Monophosphates Computed Using
  Recently Revised AMBER Parameters and Compared with Experiment. \emph{J.
  Chem. Theory Comput.} \textbf{2015}, \emph{11}, 2315--2328\relax
\mciteBstWouldAddEndPuncttrue
\mciteSetBstMidEndSepPunct{\mcitedefaultmidpunct}
{\mcitedefaultendpunct}{\mcitedefaultseppunct}\relax
\EndOfBibitem
\bibitem[Tso \latin{et~al.}(1963)Tso, Melvin, and Olson]{melvin}
Tso,~P. O.~P.; Melvin,~I.~S.; Olson,~A.~C. Interaction and Association of Bases
  and Nucleosides in Aqueous Solutions. \emph{J. Am. Chem. Soc.} \textbf{1963},
  \emph{85}, 1289--1296\relax
\mciteBstWouldAddEndPuncttrue
\mciteSetBstMidEndSepPunct{\mcitedefaultmidpunct}
{\mcitedefaultendpunct}{\mcitedefaultseppunct}\relax
\EndOfBibitem
\bibitem[Nikolova \latin{et~al.}(2013)Nikolova, Zhou, Gottardo, Alvey, Kimsey,
  and Al-Hashimi]{noncan1}
Nikolova,~E.~N.; Zhou,~H.; Gottardo,~F.~L.; Alvey,~H.~S.; Kimsey,~I.~J.;
  Al-Hashimi,~H.~M. A historical account of Hoogsteen base-pairs in duplex DNA.
  \emph{Biopolymers} \textbf{2013}, \emph{99}, 955--968\relax
\mciteBstWouldAddEndPuncttrue
\mciteSetBstMidEndSepPunct{\mcitedefaultmidpunct}
{\mcitedefaultendpunct}{\mcitedefaultseppunct}\relax
\EndOfBibitem
\bibitem[Jissy and Dutta(2014)Jissy, and Dutta]{noncan2}
Jissy,~A.~K.; Dutta,~A. Design and Applications of Noncanonical DNA Base Pairs.
  \emph{J. Phys. Chem. Lett.} \textbf{2014}, \emph{5}, 154--166\relax
\mciteBstWouldAddEndPuncttrue
\mciteSetBstMidEndSepPunct{\mcitedefaultmidpunct}
{\mcitedefaultendpunct}{\mcitedefaultseppunct}\relax
\EndOfBibitem
\bibitem[Sharp and Honig(1990)Sharp, and Honig]{honig}
Sharp,~K.~A.; Honig,~B. Calculating total electrostatic energies with the
  nonlinear Poisson-Boltzmann equation. \emph{J. Phys. Chem.} \textbf{1990},
  \emph{94}, 7684--7692\relax
\mciteBstWouldAddEndPuncttrue
\mciteSetBstMidEndSepPunct{\mcitedefaultmidpunct}
{\mcitedefaultendpunct}{\mcitedefaultseppunct}\relax
\EndOfBibitem
\bibitem[Manning(1969)]{manning}
Manning,~G.~S. Limiting Laws and Counterion Condensation in Polyelectrolyte
  Solutions I. Colligative Properties. \emph{J. Chem. Phys.} \textbf{1969},
  \emph{51}, 924--933\relax
\mciteBstWouldAddEndPuncttrue
\mciteSetBstMidEndSepPunct{\mcitedefaultmidpunct}
{\mcitedefaultendpunct}{\mcitedefaultseppunct}\relax
\EndOfBibitem
\bibitem[Olson and Manning(1976)Olson, and Manning]{olson_length}
Olson,~W.~K.; Manning,~G.~S. A configurational interpretation of the axial
  phosphate spacing in polynucleotide helices and random coils.
  \emph{Biopolymers} \textbf{1976}, \emph{15}, 2391\relax
\mciteBstWouldAddEndPuncttrue
\mciteSetBstMidEndSepPunct{\mcitedefaultmidpunct}
{\mcitedefaultendpunct}{\mcitedefaultseppunct}\relax
\EndOfBibitem
\bibitem[Hasted(1972)]{dielectric_book}
Hasted,~J.~B. \emph{Liquid water: dielectric properties; Water, a Comprehensive
  Treatise}; Plenum Press, New York, 1972\relax
\mciteBstWouldAddEndPuncttrue
\mciteSetBstMidEndSepPunct{\mcitedefaultmidpunct}
{\mcitedefaultendpunct}{\mcitedefaultseppunct}\relax
\EndOfBibitem
\bibitem[Doi and Edwards(1986)Doi, and Edwards]{doi_edwards}
Doi,~M.; Edwards,~S.~F. \emph{The Theory of Polymer Dynamics}; Clarendon Press,
  Oxford, 1986\relax
\mciteBstWouldAddEndPuncttrue
\mciteSetBstMidEndSepPunct{\mcitedefaultmidpunct}
{\mcitedefaultendpunct}{\mcitedefaultseppunct}\relax
\EndOfBibitem
\bibitem[Brunet \latin{et~al.}(2015)Brunet, Turdin, Salome, Rousseau,
  Destainville, and Manghi]{persistence_length_expt}
Brunet,~A.; Turdin,~C.; Salome,~L.; Rousseau,~P.; Destainville,~N.; Manghi,~M.
  Dependence of DNA Persistence Length on Ionic Strength of Solutions with
  Monovalent and Divalent Salts: A Joint Theory–Experiment Study.
  \emph{Macromolecules} \textbf{2015}, \emph{48}, 3641--3652\relax
\mciteBstWouldAddEndPuncttrue
\mciteSetBstMidEndSepPunct{\mcitedefaultmidpunct}
{\mcitedefaultendpunct}{\mcitedefaultseppunct}\relax
\EndOfBibitem
\bibitem[Ha and Thirumalai(1995)Ha, and Thirumalai]{Ha1995}
Ha,~B.~Y.; Thirumalai,~D. {Electrostatic Persistence Length of a
  Polyelectrolyte Chain}. \emph{Macromolecules} \textbf{1995}, \emph{28},
  577--581\relax
\mciteBstWouldAddEndPuncttrue
\mciteSetBstMidEndSepPunct{\mcitedefaultmidpunct}
{\mcitedefaultendpunct}{\mcitedefaultseppunct}\relax
\EndOfBibitem
\bibitem[Odijk(1977)]{osf_1}
Odijk,~T.~J. Polyelectrolytes near the rod limit. \emph{Polym. Sci.}
  \textbf{1977}, \emph{15}, 477\relax
\mciteBstWouldAddEndPuncttrue
\mciteSetBstMidEndSepPunct{\mcitedefaultmidpunct}
{\mcitedefaultendpunct}{\mcitedefaultseppunct}\relax
\EndOfBibitem
\bibitem[Skolnick and Fixman(1977)Skolnick, and Fixman]{osf_2}
Skolnick,~J.; Fixman,~M. Electrostatic Persistence Length of a Wormlike
  Polyelectrolyte. \emph{Macromolecules} \textbf{1977}, \emph{10}, 944\relax
\mciteBstWouldAddEndPuncttrue
\mciteSetBstMidEndSepPunct{\mcitedefaultmidpunct}
{\mcitedefaultendpunct}{\mcitedefaultseppunct}\relax
\EndOfBibitem
\bibitem[Ha and Thirumalai(1999)Ha, and Thirumalai]{esp_Ha}
Ha,~B.~Y.; Thirumalai,~D. Persistence length of flexible polyelectrolyte
  chains. \emph{J. Chem. Phys.} \textbf{1999}, \emph{110}, 7533\relax
\mciteBstWouldAddEndPuncttrue
\mciteSetBstMidEndSepPunct{\mcitedefaultmidpunct}
{\mcitedefaultendpunct}{\mcitedefaultseppunct}\relax
\EndOfBibitem
\bibitem[Netz and Orland(1999)Netz, and Orland]{esp_Netz}
Netz,~R.; Orland,~H. Variational theory for a single polyelectrolyte chain.
  \emph{Eur. Phys. J. B} \textbf{1999}, \emph{8}, 81\relax
\mciteBstWouldAddEndPuncttrue
\mciteSetBstMidEndSepPunct{\mcitedefaultmidpunct}
{\mcitedefaultendpunct}{\mcitedefaultseppunct}\relax
\EndOfBibitem
\bibitem[Honeycutt and Thirumalai(1992)Honeycutt, and Thirumalai]{honeycutt_dt}
Honeycutt,~J.~D.; Thirumalai,~D. The nature of folded states of globular
  proteins. \emph{Biopolymers} \textbf{1992}, \emph{32}, 695--709\relax
\mciteBstWouldAddEndPuncttrue
\mciteSetBstMidEndSepPunct{\mcitedefaultmidpunct}
{\mcitedefaultendpunct}{\mcitedefaultseppunct}\relax
\EndOfBibitem
\bibitem[Murphy \latin{et~al.}(2004)Murphy, Rasnik, Cheng, Lohman, and
  Ha]{Murphy2004}
Murphy,~M.; Rasnik,~I.; Cheng,~W.; Lohman,~T.; Ha,~T. {Probing single stranded
  DNA conformational flexibility using fluorescence spectroscopy}.
  \emph{Biophys. J.} \textbf{2004}, \emph{86}, 2530--2537\relax
\mciteBstWouldAddEndPuncttrue
\mciteSetBstMidEndSepPunct{\mcitedefaultmidpunct}
{\mcitedefaultendpunct}{\mcitedefaultseppunct}\relax
\EndOfBibitem
\bibitem[Kuznetsov \latin{et~al.}(2001)Kuznetsov, Shen, Benight, and
  Ansari]{Kuznetsov2001}
Kuznetsov,~S.~V.; Shen,~Y.; Benight,~A.~S.; Ansari,~A. {A semiflexible polymer
  model applied to loop formation in DNA hairpins.} \emph{Biophys. J.}
  \textbf{2001}, \emph{81}, 2864--2875\relax
\mciteBstWouldAddEndPuncttrue
\mciteSetBstMidEndSepPunct{\mcitedefaultmidpunct}
{\mcitedefaultendpunct}{\mcitedefaultseppunct}\relax
\EndOfBibitem
\bibitem[Doose \latin{et~al.}(2007)Doose, Barsch, and Sauer]{Doose2007}
Doose,~S.~S.; Barsch,~H.; Sauer,~M. {Polymer properties of polythymine as
  revealed by translational diffusion}. \emph{Biophys. J.} \textbf{2007},
  \emph{93}, 1224--1234\relax
\mciteBstWouldAddEndPuncttrue
\mciteSetBstMidEndSepPunct{\mcitedefaultmidpunct}
{\mcitedefaultendpunct}{\mcitedefaultseppunct}\relax
\EndOfBibitem
\bibitem[Chen \latin{et~al.}(2012)Chen, Meisburger, Pabit, Sutton, Webb, and
  Pollack]{Chen2012}
Chen,~H.; Meisburger,~S.~P.; Pabit,~S.~A.; Sutton,~J.~L.; Webb,~W.~W.;
  Pollack,~L. {Ionic strength-dependent persistence lengths of single-stranded
  RNA and DNA.} \emph{Proc. Natl. Acad. Sci. U. S. A.} \textbf{2012},
  \emph{109}, 799--804\relax
\mciteBstWouldAddEndPuncttrue
\mciteSetBstMidEndSepPunct{\mcitedefaultmidpunct}
{\mcitedefaultendpunct}{\mcitedefaultseppunct}\relax
\EndOfBibitem
\bibitem[Shen \latin{et~al.}(2001)Shen, Kuznetsov, and Ansari]{ansari_hairpin}
Shen,~Y.; Kuznetsov,~S.~V.; Ansari,~A. Loop Dependence of the Dynamics of DNA
  Hairpins. \emph{J. Phys. Chem. B} \textbf{2001}, \emph{105},
  12202--12211\relax
\mciteBstWouldAddEndPuncttrue
\mciteSetBstMidEndSepPunct{\mcitedefaultmidpunct}
{\mcitedefaultendpunct}{\mcitedefaultseppunct}\relax
\EndOfBibitem
\bibitem[Kuznetsov and Ansari(2012)Kuznetsov, and Ansari]{ansari_hairpin2}
Kuznetsov,~S.~V.; Ansari,~A. A Kinetic Zipper Model with Intrachain
  Interactions Applied to Nucleic Acid Hairpin Folding Kinetics. \emph{Biophys.
  J.} \textbf{2012}, \emph{102}, 101--111\relax
\mciteBstWouldAddEndPuncttrue
\mciteSetBstMidEndSepPunct{\mcitedefaultmidpunct}
{\mcitedefaultendpunct}{\mcitedefaultseppunct}\relax
\EndOfBibitem
\bibitem[Linak and Dorfman(2010)Linak, and Dorfman]{dorfman1}
Linak,~M.~C.; Dorfman,~K.~D. Analysis of a DNA simulation model through hairpin
  melting experiments. \emph{J. Chem. Phys.} \textbf{2010}, \emph{133},
  125101\relax
\mciteBstWouldAddEndPuncttrue
\mciteSetBstMidEndSepPunct{\mcitedefaultmidpunct}
{\mcitedefaultendpunct}{\mcitedefaultseppunct}\relax
\EndOfBibitem
\bibitem[Linak \latin{et~al.}(2011)Linak, Tourdot, and Dorfman]{dorfman2}
Linak,~M.~C.; Tourdot,~R.; Dorfman,~K.~D. Moving beyond Watson Crick models of
  coarse grained DNA dynamics. \emph{J. Chem. Phys.} \textbf{2011}, \emph{135},
  205102\relax
\mciteBstWouldAddEndPuncttrue
\mciteSetBstMidEndSepPunct{\mcitedefaultmidpunct}
{\mcitedefaultendpunct}{\mcitedefaultseppunct}\relax
\EndOfBibitem
\bibitem[{de Gennes}(1979)]{de_gennes}
{de Gennes},~P.~G. \emph{Scaling Concepts in Polymer Physics}; Cornell
  University Press, Ithaca, 1979\relax
\mciteBstWouldAddEndPuncttrue
\mciteSetBstMidEndSepPunct{\mcitedefaultmidpunct}
{\mcitedefaultendpunct}{\mcitedefaultseppunct}\relax
\EndOfBibitem
\bibitem[Guillou and Zinn-Justin(1977)Guillou, and Zinn-Justin]{renormal}
Guillou,~J. C.~L.; Zinn-Justin,~J. Critical Exponents for the n-Vector Model in
  Three Dimensions from Field Theory. \emph{Phys. Rev. Lett.} \textbf{1977},
  \emph{39}, 95--98\relax
\mciteBstWouldAddEndPuncttrue
\mciteSetBstMidEndSepPunct{\mcitedefaultmidpunct}
{\mcitedefaultendpunct}{\mcitedefaultseppunct}\relax
\EndOfBibitem
\bibitem[Sim \latin{et~al.}(2012)Sim, Lipfert, Herschlag, and Doniach]{Sim2012}
Sim,~A. Y.~L.; Lipfert,~J.; Herschlag,~D.; Doniach,~S. {Salt dependence of the
  radius of gyration and flexibility of single-stranded DNA in solution probed
  by small-angle x-ray scattering}. \emph{Phys. Rev. E - Stat. Nonlinear, Soft
  Matter Phys.} \textbf{2012}, \emph{82}, 021901\relax
\mciteBstWouldAddEndPuncttrue
\mciteSetBstMidEndSepPunct{\mcitedefaultmidpunct}
{\mcitedefaultendpunct}{\mcitedefaultseppunct}\relax
\EndOfBibitem
\bibitem[Guy \latin{et~al.}(2012)Guy, Piggot, and Khalid]{ssDNA_aa}
Guy,~A.~T.; Piggot,~T.~J.; Khalid,~S. Single-stranded DNA within nanopores:
  conformational dynamics and implications for sequencing; a molecular dynamics
  simulation study. \emph{Biophys. J.} \textbf{2012}, \emph{103},
  1028--1036\relax
\mciteBstWouldAddEndPuncttrue
\mciteSetBstMidEndSepPunct{\mcitedefaultmidpunct}
{\mcitedefaultendpunct}{\mcitedefaultseppunct}\relax
\EndOfBibitem
\bibitem[Plumridge \latin{et~al.}(2017)Plumridge, Meisburger, Andersen, and
  Pollack]{pollack_saxs}
Plumridge,~A.; Meisburger,~S.~P.; Andersen,~K.; Pollack,~L. Visualizing
  single-stranded nucleic acids in solution. \emph{Nucleic Acid Res.}
  \textbf{2017}, \emph{45}, 3932--3943\relax
\mciteBstWouldAddEndPuncttrue
\mciteSetBstMidEndSepPunct{\mcitedefaultmidpunct}
{\mcitedefaultendpunct}{\mcitedefaultseppunct}\relax
\EndOfBibitem
\bibitem[Isaksson \latin{et~al.}(2004)Isaksson, Acharya, Barman, Cheruku, and
  Chattopadhyaya]{base_stacking_dA}
Isaksson,~J.; Acharya,~S.; Barman,~J.; Cheruku,~P.; Chattopadhyaya,~J.
  Single-Stranded Adenine-Rich DNA and RNA Retain Structural Characteristics of
  Their Respective Double-Stranded Conformations and Show Directional
  Differences in Stacking Pattern. \emph{Biochemistry} \textbf{2004},
  \emph{43}, 15996--16010\relax
\mciteBstWouldAddEndPuncttrue
\mciteSetBstMidEndSepPunct{\mcitedefaultmidpunct}
{\mcitedefaultendpunct}{\mcitedefaultseppunct}\relax
\EndOfBibitem
\bibitem[Ke \latin{et~al.}(2007)Ke, Humeniuk, S-Gracz, and Marszalek]{fex1}
Ke,~C.; Humeniuk,~M.; S-Gracz,~H.; Marszalek,~P.~E. Direct Measurements of Base
  Stacking Interactions in DNA by Single-Molecule Atomic-Force Spectroscopy.
  \emph{Phys. Rev. Lett.} \textbf{2007}, \emph{99}, 018302\relax
\mciteBstWouldAddEndPuncttrue
\mciteSetBstMidEndSepPunct{\mcitedefaultmidpunct}
{\mcitedefaultendpunct}{\mcitedefaultseppunct}\relax
\EndOfBibitem
\bibitem[Kohn \latin{et~al.}(2004)Kohn, Millett, Jacob, Zagrovic, Dillon,
  Cingel, Dothager, Seifert, Thiyagarajan, Sosnick, Hasan, Pande, Ruczinski,
  Doniach, and Plaxco]{Kohn2004}
Kohn,~J.~E.; Millett,~I.~S.; Jacob,~J.; Zagrovic,~B.; Dillon,~T.~M.;
  Cingel,~N.; Dothager,~R.~S.; Seifert,~S.; Thiyagarajan,~P.; Sosnick,~T.~R.;
  Hasan,~M.~Z.; Pande,~V.~S.; Ruczinski,~I.; Doniach,~S.; Plaxco,~K.~W.
  {Random-coil behavior and the dimensions of chemically unfolded proteins}.
  \emph{Proc. Natl. Acad. Sci. U. S. A.} \textbf{2004}, \emph{101},
  12491--12496\relax
\mciteBstWouldAddEndPuncttrue
\mciteSetBstMidEndSepPunct{\mcitedefaultmidpunct}
{\mcitedefaultendpunct}{\mcitedefaultseppunct}\relax
\EndOfBibitem
\bibitem[Wilkins \latin{et~al.}(1999)Wilkins, Grimshaw, Receveur, Dobson,
  Jones, and Smith]{Wilkins99}
Wilkins,~D.~K.; Grimshaw,~S.~B.; Receveur,~V.; Dobson,~C.~M.; Jones,~J.~A.;
  Smith,~L.~J. {Hydrodynamic Radii of Native and Denatured Proteins Measured by
  Pulse Field Gradient NMR Techniques}. \emph{Biochemistry} \textbf{1999},
  \emph{38}, 16424--16431\relax
\mciteBstWouldAddEndPuncttrue
\mciteSetBstMidEndSepPunct{\mcitedefaultmidpunct}
{\mcitedefaultendpunct}{\mcitedefaultseppunct}\relax
\EndOfBibitem
\bibitem[Tinland \latin{et~al.}(1997)Tinland, Pluen, Sturm, and
  Weill]{tinland_ssDNA}
Tinland,~B.; Pluen,~A.; Sturm,~J.; Weill,~G. Persistence Length of
  Single-Stranded DNA. \emph{Macromolecules} \textbf{1997}, \emph{30},
  5763--5765\relax
\mciteBstWouldAddEndPuncttrue
\mciteSetBstMidEndSepPunct{\mcitedefaultmidpunct}
{\mcitedefaultendpunct}{\mcitedefaultseppunct}\relax
\EndOfBibitem
\bibitem[Gubarev \latin{et~al.}(2009)Gubarev, Carrillo, and
  Dobrynin]{biexp_fit}
Gubarev,~A.; Carrillo,~J.-M.~Y.; Dobrynin,~A.~V. Scale-Dependent Electrostatic
  Stiffening in Biopolymers. \emph{Macromolecules} \textbf{2009}, \emph{42},
  5851--5860\relax
\mciteBstWouldAddEndPuncttrue
\mciteSetBstMidEndSepPunct{\mcitedefaultmidpunct}
{\mcitedefaultendpunct}{\mcitedefaultseppunct}\relax
\EndOfBibitem
\bibitem[Toan and Thirumalai(2012)Toan, and Thirumalai]{Toan2012}
Toan,~N.~M.; Thirumalai,~D. {On the origin of the unusual behavior in the
  stretching of single-stranded DNA}. \emph{J. Chem. Phys.} \textbf{2012},
  \emph{136}, 235103\relax
\mciteBstWouldAddEndPuncttrue
\mciteSetBstMidEndSepPunct{\mcitedefaultmidpunct}
{\mcitedefaultendpunct}{\mcitedefaultseppunct}\relax
\EndOfBibitem
\bibitem[Knotts \latin{et~al.}(2007)Knotts, Rathore, Schwartz, and
  de~Pablo]{depablo_3spn0}
Knotts,~T.~A.; Rathore,~N.; Schwartz,~D.~C.; de~Pablo,~J.~J. A coarse-grained
  model for DNA. \emph{J. Chem. Phys.} \textbf{2007}, \emph{126}, 084901\relax
\mciteBstWouldAddEndPuncttrue
\mciteSetBstMidEndSepPunct{\mcitedefaultmidpunct}
{\mcitedefaultendpunct}{\mcitedefaultseppunct}\relax
\EndOfBibitem
\bibitem[Freeman \latin{et~al.}(2014)Freeman, Hinckley, Lequieu, Whitmer, and
  de~Pablo]{depablo_3spn2c}
Freeman,~G.~S.; Hinckley,~D.~M.; Lequieu,~J.~P.; Whitmer,~J.~K.;
  de~Pablo,~J.~J. Coarse-grained modeling of DNA curvature. \emph{J. Chem.
  Phys.} \textbf{2014}, \emph{141}, 165103\relax
\mciteBstWouldAddEndPuncttrue
\mciteSetBstMidEndSepPunct{\mcitedefaultmidpunct}
{\mcitedefaultendpunct}{\mcitedefaultseppunct}\relax
\EndOfBibitem
\bibitem[McIntosh \latin{et~al.}(2009)McIntosh, Ribeck, and Saleh]{bare_ssDNA}
McIntosh,~D.~B.; Ribeck,~N.; Saleh,~O.~A. Detailed scaling analysis of
  low-force polyelectrolyte elasticity. \emph{Phys. Rev. E} \textbf{2009},
  \emph{80}, 041803\relax
\mciteBstWouldAddEndPuncttrue
\mciteSetBstMidEndSepPunct{\mcitedefaultmidpunct}
{\mcitedefaultendpunct}{\mcitedefaultseppunct}\relax
\EndOfBibitem
\bibitem[Jacobson \latin{et~al.}(2017)Jacobson, McIntosh, Stevens, Runinstein,
  and Saleh]{bare_ssDNA2}
Jacobson,~D.~R.; McIntosh,~D.~B.; Stevens,~M.~J.; Runinstein,~M.; Saleh,~O.~A.
  Single-stranded nucleic acid elasticity arises from internal electrostatic
  tension. \emph{Proc. Natl. Acad. Sci.} \textbf{2017}, \emph{114},
  5095--5100\relax
\mciteBstWouldAddEndPuncttrue
\mciteSetBstMidEndSepPunct{\mcitedefaultmidpunct}
{\mcitedefaultendpunct}{\mcitedefaultseppunct}\relax
\EndOfBibitem
\bibitem[Goddard \latin{et~al.}(2000)Goddard, Bonnet, Krichevsky, and
  Libchaber]{Goddard}
Goddard,~N.~L.; Bonnet,~G.; Krichevsky,~O.; Libchaber,~A. Sequence dependent
  rigidity of single stranded DNA. \emph{Phys. Rev. Lett.} \textbf{2000},
  \emph{85}, 2400\relax
\mciteBstWouldAddEndPuncttrue
\mciteSetBstMidEndSepPunct{\mcitedefaultmidpunct}
{\mcitedefaultendpunct}{\mcitedefaultseppunct}\relax
\EndOfBibitem
\bibitem[Seol \latin{et~al.}(2007)Seol, Skinner, and Visscher]{fex2}
Seol,~Y.; Skinner,~G.~M.; Visscher,~K. Stretching of Homopolymeric RNA Reveals
  Single-Stranded Helices and Base-Stacking. \emph{Phys. Rev. Lett.}
  \textbf{2007}, \emph{98}, 158103\relax
\mciteBstWouldAddEndPuncttrue
\mciteSetBstMidEndSepPunct{\mcitedefaultmidpunct}
{\mcitedefaultendpunct}{\mcitedefaultseppunct}\relax
\EndOfBibitem
\bibitem[McIntosh \latin{et~al.}(2014)McIntosh, Duggan, Gouil, and Saleh]{fex3}
McIntosh,~D.~B.; Duggan,~G.; Gouil,~Q.; Saleh,~O.~A. Sequence-dependent
  elasticity and electrostatics of single-stranded DNA: signatures of
  base-stacking. \emph{Biophys. J.} \textbf{2014}, \emph{106}, 659--666\relax
\mciteBstWouldAddEndPuncttrue
\mciteSetBstMidEndSepPunct{\mcitedefaultmidpunct}
{\mcitedefaultendpunct}{\mcitedefaultseppunct}\relax
\EndOfBibitem
\bibitem[Marko and Siggia(1995)Marko, and Siggia]{marko_siggia}
Marko,~J.~E.; Siggia,~E.~D. Stretching DNA. \emph{Macromolecules}
  \textbf{1995}, \emph{28}, 8759--8770\relax
\mciteBstWouldAddEndPuncttrue
\mciteSetBstMidEndSepPunct{\mcitedefaultmidpunct}
{\mcitedefaultendpunct}{\mcitedefaultseppunct}\relax
\EndOfBibitem
\bibitem[Buhot and Halperin(2004)Buhot, and Halperin]{halperin}
Buhot,~A.; Halperin,~A. Effects of stacking on the configurations and
  elasticity of single-stranded nucleic acids. \emph{Phys. Rev. E.}
  \textbf{2004}, \emph{70}, 020902\relax
\mciteBstWouldAddEndPuncttrue
\mciteSetBstMidEndSepPunct{\mcitedefaultmidpunct}
{\mcitedefaultendpunct}{\mcitedefaultseppunct}\relax
\EndOfBibitem
\bibitem[Holbrook \latin{et~al.}(1999)Holbrook, Capp, Saecker, and
  Record]{holbrok}
Holbrook,~J.~A.; Capp,~M.~W.; Saecker,~R.~M.; Record,~M.~T. Enthalpy and Heat
  Capacity Changes for Formation of an Oligomeric DNA Duplex:  Interpretation
  in Terms of Coupled Processes of Formation and Association of Single-Stranded
  Helices. \emph{Biochemistry} \textbf{1999}, \emph{38}, 8409--8422\relax
\mciteBstWouldAddEndPuncttrue
\mciteSetBstMidEndSepPunct{\mcitedefaultmidpunct}
{\mcitedefaultendpunct}{\mcitedefaultseppunct}\relax
\EndOfBibitem
\bibitem[Baumann \latin{et~al.}(1997)Baumann, Smith, Bloomfield, and
  Bustamante]{Baumann1997}
Baumann,~C.~G.; Smith,~S.~B.; Bloomfield,~V.~A.; Bustamante,~C. {Ionic effects
  on the elasticity of single DNA molecules}. \emph{Proc. Natl. Acad. Sci. USA}
  \textbf{1997}, \emph{94}, 6185--6190\relax
\mciteBstWouldAddEndPuncttrue
\mciteSetBstMidEndSepPunct{\mcitedefaultmidpunct}
{\mcitedefaultendpunct}{\mcitedefaultseppunct}\relax
\EndOfBibitem
\bibitem[Smith \latin{et~al.}(1996)Smith, Cui, and Bustamante]{smith_science}
Smith,~S.~B.; Cui,~Y.; Bustamante,~C. Overstretching B-DNA: The Elastic
  Response of Individual Double-Stranded and Single-Stranded DNA Molecules.
  \emph{Science} \textbf{1996}, \emph{271}, 795--799\relax
\mciteBstWouldAddEndPuncttrue
\mciteSetBstMidEndSepPunct{\mcitedefaultmidpunct}
{\mcitedefaultendpunct}{\mcitedefaultseppunct}\relax
\EndOfBibitem
\bibitem[Harrington(1978)]{harrington_lp}
Harrington,~R.~E. Opticohydrodynamic properties of high-molecular-weight DNA.
  III. The effects of NaCl concentration. \emph{Biopolymers} \textbf{1978},
  \emph{17}, 919--936\relax
\mciteBstWouldAddEndPuncttrue
\mciteSetBstMidEndSepPunct{\mcitedefaultmidpunct}
{\mcitedefaultendpunct}{\mcitedefaultseppunct}\relax
\EndOfBibitem
\bibitem[Sobel and Harpst(1991)Sobel, and Harpst]{sobel_lp}
Sobel,~E.~S.; Harpst,~J.~A. Effect of Na$^{+}$ on the persistence length and
  excluded volume of T7 bacteriophage DNA. \emph{Biopolymers} \textbf{1991},
  \emph{31}, 1559--1564\relax
\mciteBstWouldAddEndPuncttrue
\mciteSetBstMidEndSepPunct{\mcitedefaultmidpunct}
{\mcitedefaultendpunct}{\mcitedefaultseppunct}\relax
\EndOfBibitem
\bibitem[Maret and Weill(1983)Maret, and Weill]{maret_lp}
Maret,~G.; Weill,~G. Magnetic birefringence study of the electrostatic and
  intrinsic persistence length of DNA. \emph{Biopolymers} \textbf{1983},
  \emph{22}, 2727--2744\relax
\mciteBstWouldAddEndPuncttrue
\mciteSetBstMidEndSepPunct{\mcitedefaultmidpunct}
{\mcitedefaultendpunct}{\mcitedefaultseppunct}\relax
\EndOfBibitem
\bibitem[Snodin \latin{et~al.}(2015)Snodin, Randisi, Mosayebi, {\v{S}}ulc,
  Schreck, Romano, Ouldridge, Tsukanov, Nir, Louis, and Doye]{oxDNA2}
Snodin,~B. E.~K.; Randisi,~F.; Mosayebi,~M.; {\v{S}}ulc,~P.; Schreck,~J.~S.;
  Romano,~F.; Ouldridge,~T.~E.; Tsukanov,~R.; Nir,~E.; Louis,~A.~A.; Doye,~J.
  P.~K. {Introducing improved structural properties and salt dependence into a
  coarse-grained model of DNA}. \emph{J. Chem. Phys.} \textbf{2015},
  \emph{142}\relax
\mciteBstWouldAddEndPuncttrue
\mciteSetBstMidEndSepPunct{\mcitedefaultmidpunct}
{\mcitedefaultendpunct}{\mcitedefaultseppunct}\relax
\EndOfBibitem
\bibitem[Mitchell \latin{et~al.}(2017)Mitchell, Glowacki, Grandchamp, Manning,
  and Maddocks]{maddocks}
Mitchell,~J.~S.; Glowacki,~J.; Grandchamp,~A.~E.; Manning,~R.~S.;
  Maddocks,~J.~H. {Sequence-Dependent Persistence Lengths of DNA}. \emph{J.
  Chem. Theory Comput.} \textbf{2017}, \emph{13}, 1539--1555\relax
\mciteBstWouldAddEndPuncttrue
\mciteSetBstMidEndSepPunct{\mcitedefaultmidpunct}
{\mcitedefaultendpunct}{\mcitedefaultseppunct}\relax
\EndOfBibitem
\bibitem[Geggier and Vologodskii(2010)Geggier, and Vologodskii]{volgod}
Geggier,~S.; Vologodskii,~A. Sequence dependence of DNA bending rigidity.
  \emph{Proc. Natl. Acad. Sci. USA} \textbf{2010}, \emph{107},
  15421--15426\relax
\mciteBstWouldAddEndPuncttrue
\mciteSetBstMidEndSepPunct{\mcitedefaultmidpunct}
{\mcitedefaultendpunct}{\mcitedefaultseppunct}\relax
\EndOfBibitem
\bibitem[Denesyuk and Thirumalai(2015)Denesyuk, and
  Thirumalai]{Denesyuk_nature}
Denesyuk,~N.; Thirumalai,~D. How do metal ions direct ribozyme folding?
  \emph{Nat. Chem.} \textbf{2015}, \emph{7}, 793--801\relax
\mciteBstWouldAddEndPuncttrue
\mciteSetBstMidEndSepPunct{\mcitedefaultmidpunct}
{\mcitedefaultendpunct}{\mcitedefaultseppunct}\relax
\EndOfBibitem
\end{mcitethebibliography}

\providecommand{\latin}[1]{#1}
\providecommand*\mcitethebibliography{\thebibliography}
\csname @ifundefined\endcsname{endmcitethebibliography}
  {\let\endmcitethebibliography\endthebibliography}{}

\end{document}